\documentclass[prc,a4paper,showpacs,superscriptaddress]{revtex4}

\usepackage{graphicx}

\begin{document}

\title{Spallation Neutron Production by 0.8, 1.2 and 
1.6~GeV Protons on various Targets.}

\author{S.~Leray}
\email{sleray@cea.fr} 
\affiliation{CEA/Saclay, DAPNIA/SPhN, 91191 Gif-sur-Yvette Cedex, France}
\affiliation{Laboratoire National SATURNE, Saclay, France}

\author{F.~Borne}
\affiliation{CEA/DIF/DPTA, Bruy\`eres-le-Ch\^atel, France}

\author{A.~Boudard}
\affiliation{CEA/Saclay, DAPNIA/SPhN, 91191 Gif-sur-Yvette Cedex, France}

\author{F.~Brochard}
\affiliation{Laboratoire National SATURNE, Saclay, France} 

\author{S.~Crespin}
\affiliation{CEA/DIF/DPTA, Bruy\`eres-le-Ch\^atel, France}

\author{D.~Drake}
\affiliation{Laboratoire National SATURNE, Saclay, France} 

\author{J.C.~Duchazeaubeneix}
\affiliation{Laboratoire National SATURNE, Saclay, France} 

\author{D.~Durand}
\affiliation{LPC, Caen, France}

\author{J.M.~Durand}
\affiliation{Laboratoire National SATURNE, Saclay, France} 

\author{J.~Fr\'ehaut}
\affiliation{CEA/DIF/DPTA, Bruy\`eres-le-Ch\^atel, France}

\author{F.~Hanappe}
\affiliation{Universit\'e Libre de Bruxelles, Belgium}

\author{C.~Le Brun}
\affiliation{LPC, Caen, France}

\author{F.R.~Lecolley}
\affiliation{LPC, Caen, France}

\author{J.F.~Lecolley}
\affiliation{LPC, Caen, France}

\author{X.~Ledoux}
\affiliation{CEA/DIF/DPTA, Bruy\`eres-le-Ch\^atel, France}

\author{F.~Lefebvres}
\affiliation{LPC, Caen, France}

\author{R.~Legrain}
\affiliation{CEA/Saclay, DAPNIA/SPhN, 91191 Gif-sur-Yvette Cedex, France}

\author{M.~Louvel}
\affiliation{LPC, Caen, France}

\author{E.~Martinez}
\affiliation{CEA/DIF/DPTA, Bruy\`eres-le-Ch\^atel, France}

\author{S.I.~Meigo}
\affiliation{Laboratoire National SATURNE, Saclay, France} 

\author{S.~M\'enard}
\affiliation{IPN, Orsay, France}

\author{G.~Milleret}
\affiliation{Laboratoire National SATURNE, Saclay, France} 

\author{Y.~Patin}
\affiliation{CEA/DIF/DPTA, Bruy\`eres-le-Ch\^atel, France}

\author{E.~Petibon}
\affiliation{CEA/DIF/DPTA, Bruy\`eres-le-Ch\^atel, France}

\author{P.~Pras}
\affiliation{CEA/DIF/DPTA, Bruy\`eres-le-Ch\^atel, France}

\author{L.~Stuttge}
\affiliation{IReS, Strasbourg, France}

\author{Y.~Terrien}
\affiliation{CEA/Saclay, DAPNIA/SPhN, 91191 Gif-sur-Yvette Cedex, France}

\author{J.~Thun}
\affiliation{Laboratoire National SATURNE, Saclay, France} 
\affiliation{Uppsala University, Sweden}

\author{C.~Varignon}
\affiliation{LPC, Caen, France}

\author{D.M.~Whittal}
\affiliation{Laboratoire National SATURNE, Saclay, France} 

\author{W.~Wlazlo}
\affiliation{Laboratoire National SATURNE, Saclay, France} 

\date{February 6, 2002}

\begin{abstract}

Spallation neutron production in proton induced reactions on Al, Fe, Zr, W, Pb and Th targets at 1.2 GeV 
and on Fe and Pb at 0.8, and 1.6 GeV measured at the SATURNE accelerator in Saclay is reported. The 
experimental double-differential cross-sections are compared with calculations performed with different 
intra-nuclear cascade models implemented in high energy transport codes. The broad angular coverage 
also allowed the determination of average neutron multiplicities above 2 MeV. Deficiencies in some of the 
models 
commonly used for applications are pointed out.

\end{abstract}


\maketitle

\section{Introduction}

Large numbers of neutrons can be produced through spallation reactions induced by an intermediate energy 
(around 1 GeV) proton accelerator on a heavy element target. With the progress in high intensity 
accelerators it is now possible to conceive spallation sources that could compete with high flux reactors. 
Several spallation sources for solid state and material physics are under construction or study in the USA 
(SNS~\cite{SNS}), in Europe (SINQ~\cite{SINQ}, ESS~\cite{ESS}) and in Japan (NSP~\cite{NSP}). Spallation 
neutrons can also be used in Accelerator Driven Systems (ADS) to drive sub-critical reactors, in which 
long-lived nuclear waste could be burnt~\cite{Bow,tak} or energy produced~\cite{Rub}. All these systems 
have in common a spallation target made of a heavy, either solid (W, Pb) or liquid (Hg, Pb, Pb-Bi 
eutectic) 
metal in a container (generally steel) which is separated from the vacuum of the accelerator by a thin 
window.  

A detailed engineering design of a spallation target needs a precise optimisation of its performances in 
terms of useful neutron production and a proper assessment of specific problems likely to occur in such 
systems, like induced radioactivity, radiation damage in target, window or structure materials, additional 
required shielding due to the presence of high energy neutrons, etc... This can be done by using 
Monte-Carlo transport codes describing the interaction and transport of all the particles created in 
nuclear reactions occurring inside the system to be designed. Generally, a high energy transport code 
(often based on the HETC code from \cite{HETC}), in which elementary interactions are generated by nuclear 
physics models, is coupled below 20 MeV to a neutron transport code like MCNP~\cite{MCNP} that uses 
evaluated data files. Although the spallation mechanism has been known for many years, the models used in 
such codes, Intra-Nuclear Cascade followed by evaporation-fission, have never been really validated on 
experimental data and large discrepancies remain both between experimental data and model predictions and 
between different models. This was particurlarly obvious from the OECD/NEA 
intercomparisons~\cite{ICC1,ICC2,ICC3} of these codes, regarding neutron and residue production. This led 
to the conclusion that many improvements of the models are still needed but also that there was a lack of 
experimental data to make a good validation, especially above $800$ MeV. Among the needed data, the energy 
and angular distributions of spallation produced neutrons are essential for model probing: the high energy 
part of the spectrum allows the testing of the intra-nuclear cascade while the low energy part of the 
spectrum is sensitive to the excitation energy at the end of the intra-nuclear cascade stage and the 
evaporation model. They are also important to optimize the target geometry since secondary particles 
contribute to the propagation of the inter-nuclear cascade in a thick target and high energy neutrons are 
responsible for radiation damage in target and structural materials.

During the last years, a wide effort has been made in several laboratories to measure spallation 
data regarding neutron multiplicity distributions~\cite{Pie,Led,Lot,Hil}, light charged 
particles~\cite{NESSI} and heavy residues~\cite{Mic,Wla,TE,GSI-Au} in order to establish a base for the 
test and validation of the spallation physics models. We report in this paper on neutron production 
double-differential cross-sections, measured at the SATURNE synchrotron, induced by $0.8$, $1.2$ and $1.6$ 
GeV protons impinging on different targets. The experimental setup, although already discussed in detail 
in~\cite{Bor,Mar}, is presented in section 1 while the results are displayed in section 2. The aim was to 
measure at one energy (1.2 GeV) neutron spectra on nuclei representative of different parts of the 
periodic table of elements and at the same time corresponding to materials used in targets or structures 
of accelerator driven systems: Al, Fe, Zr, W, Pb and Th. At the two other energies only Fe and Pb targets 
were studied. Angular distributions covering $0^\circ$ to $160^\circ$ were obtained and allowed the 
determination of average neutron multiplicities per reaction above 2 MeV. The Pb data have already been 
published in a letter~\cite{XL} but are again reported here for the sake of completeness. Section 3 is 
devoted to the comparison of the data to different intra-nuclear cascade models.
 
\section{Experimental apparatus}

\begin{figure}[hbt]
	\includegraphics[width=8.0cm,angle=-90]{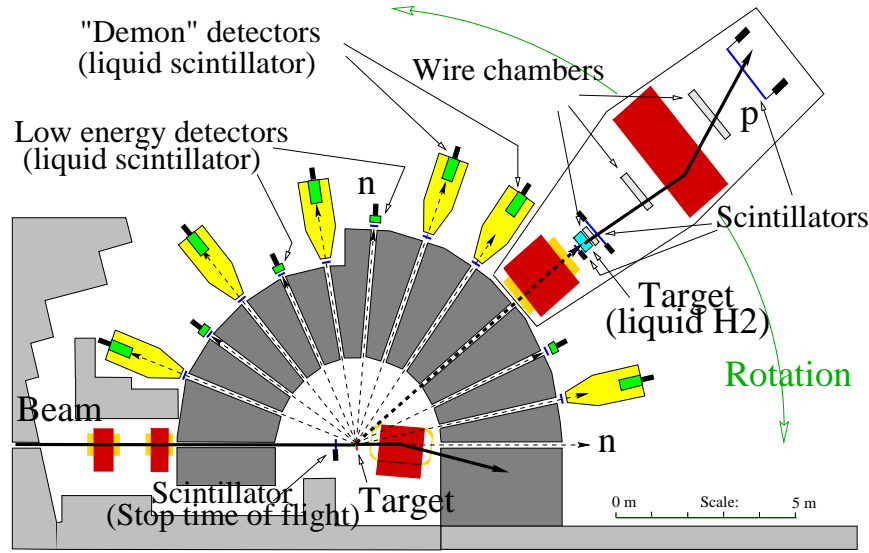}
	\caption{Experimental area with time-of-flight and spectrometer setup.}
	\label{figairexpspa}
\end{figure}

The slow extraction of the beam delivered by the Saturne synchrotron did not allow conventional 
time-of-flight measurement using the HF signal of the accelerator. Therefore, the time-of-flight had to be 
measured between the incident proton passing through a thin scintillator placed in the beam and the 
detected neutron. For the highest neutron velocities this method becomes highly imprecise due to the 
limited available flight path, and a measurement of the (n,p) charge exchange by a spectrometer technique 
was adopted. These two methods had previously been tested in a first experiment in which neutrons were 
measured at 0$^{\circ}$ and have been respectively described in detail in two previous 
papers~\cite{Bor,Mar}. Then, a new experimental area was designed to allow complete angular distribution 
measurements. The scheme of this experimental area is shown in Fig.~\ref{figairexpspa}. The beam comes 
from the left, hits the studied target and is deflected to a beam stopper (composed of lead and tungsten 
blocks) by a dipole magnet. To ensure the detection of the neutrons emitted by the target and not those 
coming from the beam stop, a large shielding of heavy concrete was built around the target. Because the 
most energetic neutrons are emitted in the forward direction this shielding is thicker between 0$^{\circ}$ 
and 90$^{\circ}$ (3.5 meters) than at backward angles (2.5 meters). The shielding is pierced by 12 
circular holes at 0$^{\circ}$ and every 15$^{\circ}$ from 10$^{\circ}$ to 160$^{\circ}$. Each channel is 
composed of two consecutive cylinders of respectively $110~mm$ and $136~mm$ in diameter with a length of 
respectively $1600~mm$ and $1800~mm$ at forward angles and $1100~mm$ and $1400~mm$ at backward angles. The 
solid angle is determined by the size of the neutron detectors, not the collimator. The two detection 
methods were used in dedicated runs due to the fact that they required different beam intensities: between 
$10^9$ and $10^{11}$ particles per second for the spectrometer (because of the low neutron-to-proton 
charge exchange efficiency) while the time-of-flight was limited to less than $10^6$ particles per second 
by the use of the in-beam scintillator. The spectrometer was not used at angles larger than 85$^{\circ}$ 
since very high energy neutrons are expected only at forward angles. 

\subsection{The time-of-flight method.}

The Saturne synchrotron was delivering continuous beam during spills of $\sim 500 ms$ with a repetition 
frequency depending on the beam energy (typically 1.5 seconds for 800~MeV protons in this experiment). A 
classic time-of-flight measurement between the neutron detector and a HF signal of the accelerator was 
thus not possible. We just recall here the principle of the method and the modifications performed since 
the first experiment at $0^{\circ}$~\cite{Bor}.

\begin{figure}[htb]
	\includegraphics[width=8.0cm,angle=-90]{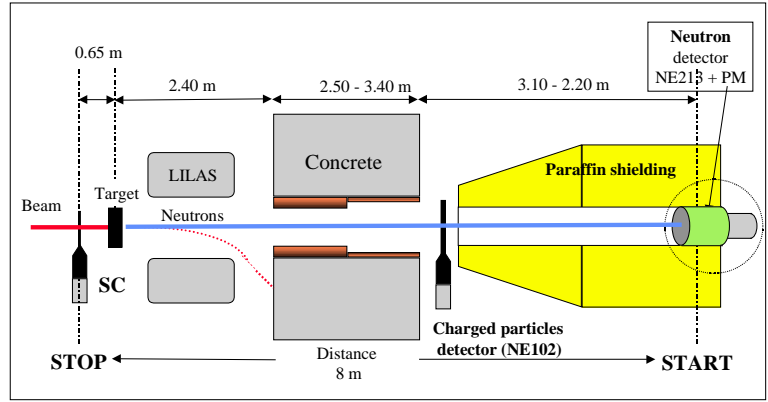}
	\caption{Principle of the time-of-flight method shown here for the 0$^{\circ}$ measurement with 	
the dipole magnet ensuring the deflection of the incident beam.}
	\label{fig:dispotof}
\end{figure}

The time-of-flight is measured between the incident proton, tagged by a thin plastic scintillator (SC), 
and a neutron sensitive NE213 liquid scintillator (see Fig~\ref{fig:dispotof}). The beam intensity is 
fixed at a maximum of 10$^{6}$ particles per second so that individual incident protons could be counted 
by SC. The target-detector distance depends on the angle but is about 8.5~meters. Up to ten angles can be 
explored simultaneously using neutron detectors composed of a cylindrical cell of NE213 liquid 
scintillator coupled to a photomultiplier. Six of them are cells of the multi-detector DEMON~\cite{Til} 
and the other four (called DENSE) are smaller detectors, optimized for low energy measurements. The latter 
detectors allow energy measurements with a reasonable precision from 2 to 14~MeV, while the DEMON cells 
are used between 4 and 400~MeV. The characteristics of the DEMON and DENSE detectors are given in 
table~\ref{table:pmcharacterstics}.

\begin{center}
	\begin{table}[htb]
	\centering
	\begin{tabular}{|c||c|c|} 
		{\bf {Characteristics}} 	& {\bf {DENSE}} 		& {\bf {DEMON}}\\ 		
\hline\hline
		{\bf {Liquid scintillator}}	& NE213				& NE213\\ \hline
		{\bf {Cell diameter}}		& 127~mm			& 160~mm\\ \hline
		{\bf {Cell length}}		& 51~mm				& 200~mm\\ \hline
		{\bf {Photomultiplier type}}	& 9390 KB			& XP 4512\\ \hline
		{\bf {Detector threshold}}	& 1.~MeV			& 1.9~MeV\\ 
	\end{tabular}
	\caption{Neutron detector characteristics.}
	\label{table:pmcharacterstics}
	\end{table}
\end{center}

The DEMON detectors are placed in a shielding of paraffin loaded with Borax and Lithium to reduce the 
background. The DENSE detectors, smaller and thus less sensitive, do not need such a protection. The 
energy threshold of the detectors is adjusted using the Compton edge of the gamma rays delivered by a 
$^{22}$Na and a $^{137}$Cs radioactive source. The detection thresholds of the DENSE and DEMON are tuned 
to 1.0 and 1.9 MeV respectively. This allows measurements with a sufficiently well defined efficiency 
above 2 and 4 MeV respectively. 

\begin{figure}[htb]
	\includegraphics[width=8.0cm,angle=-90]{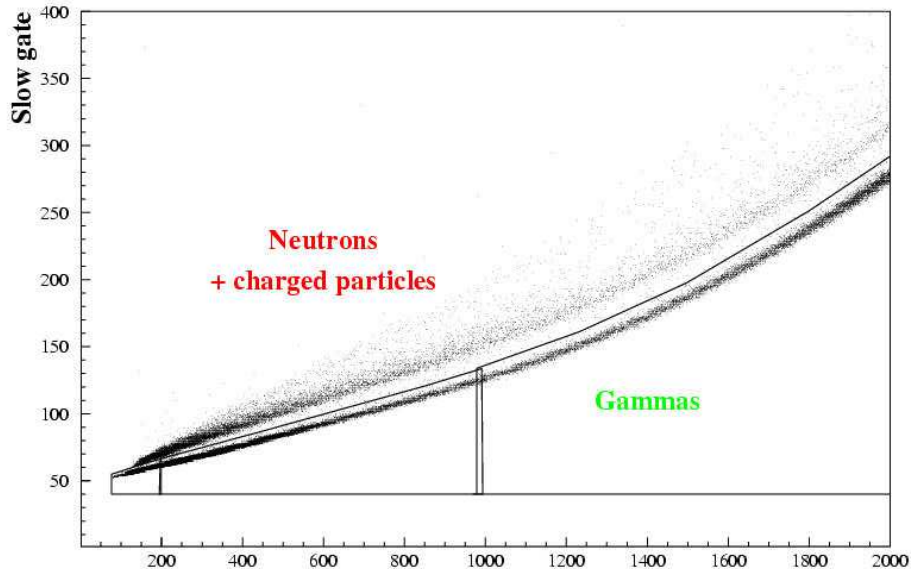}
	\caption{Neutron gamma discrimination.}
	\label{fig:discringamma}
\end{figure}

These detectors are sensitive to neutrons, $\gamma$-rays and charged particles. A plastic scintillator 
NE102 placed in front of each counter (see Fig~\ref{fig:dispotof}) tags events induced by a charged 
particle. The neutron-gamma discrimination is performed by a pulse shape analysis done as follows 
~\cite{Til}: the charge delivered by the photomultiplier is measured by a QDC 1612F during two different 
time intervals, a prompt one (125~ns long) and a delayed one (185~ns long delayed by 65~ns) giving two 
charge values (Qf) and (Qs) respectively. A bidimensional spectra Qf vs Qs allows the separation of 
neutrons and gammas (see Fig.~\ref{fig:discringamma}).

\DeclareGraphicsRule{.jpeg}{eps}{.jpeg.bb}{`convert #1 'eps:-' }

The neutron detector starts a gate of 500~ns, longer than the time-of-flight of 2~MeV neutrons (440~ns to 
cover the 8.5~meters from the target to the detector). The signal from the scintillator (SC) in the beam 
is delayed by 500~ns and is the stop of the time measurement. Due to the beam intensity, although limited 
to less than $10^6 p/s$, several protons can be detected during the gate, but only one has induced the 
detected neutron. For a common start (the neutron) up to ten stop signals are converted and stored with a 
multistop module (LECROY TDC 3377). The sum of these time spectra contains the real and accidental events. 
The accidental contribution is determined by the time measurement of unmatched start and stop signals. The 
background effect is taken into account by a measurement with an empty frame at the location of the 
target.

\begin{figure}[bth]
	\includegraphics[width=8.0cm,angle=-90]{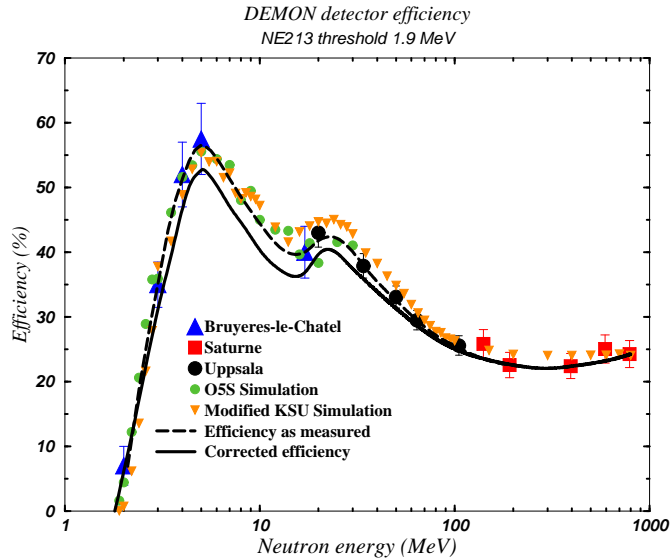} 
	\caption{Efficiency of a DEMON neutron detector as a function of energy measured at 
Bruy\`eres-le-Ch\^atel, Uppsala and Saturne (black symbols) and calculated with O5S~\protect\cite{O5S} and a 
modified version of the KSU~\protect\cite{KSU} code (grey symbols). The solid line represent the final 
parameterization used in the data analysis after correction of the neutron flux attenuation in air along 
the flight path in Saturne and in the NE102 scintillateur in front of each detector.}
	\label{fig:figeffi1}
\end{figure}

\begin{figure}[bth]
	\includegraphics[width=6.0cm,angle=-90]{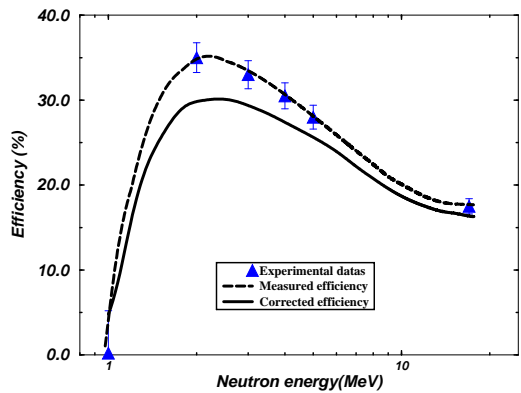} 
	\caption{Efficiency of a DENSE neutron detector as a function of energy. The symbols represent the 
values measured at Bruy\`eres-le-Ch\^atel and the dashed line a fit of these values. The solid line is the 
final parameterization used in the data analysis (after correction of the neutron flux attenuations in 
Saturne set-up).}
	\label{fig:figeffi2}
\end{figure}

The knowledge of the neutron detector efficiency being crucial for this experiment, measurements and 
calculations have been performed to determine it over the whole energy range. From 2 to 17~MeV, 
measurements were made at the Bruy\`eres-le-Ch\^atel Van de Graff accelerators as described in~\cite{Bor}. The 
quasi mono-energetic neutrons are produced by $^{7}$Li(p,n)$^{7}$Be, $^{3}$H(p,n)$^{3}$He, 
$^{3}$H(d,n)$^{4}$He and $^{2}$H(d,n)$^{3}$He reactions and the efficiency is determined by comparison 
with a standard detector (full triangles in fig.\ref{fig:figeffi1} and fig.~\ref{fig:figeffi2}). 

At higher energies (30~$\leq$~E~$\leq$~100~MeV), experiments have been performed at the TSL Uppsala 
facilities in Sweden~\cite{Thun}. A neutron beam is produced by $^{7}$Li(p,n)$^{7}$Be reaction in the 
[100-180~MeV] range. The neutron detector efficiency is then measured using n-p elastic scattering and 
the simultaneous detection of the correlated n and p; the so called associated particle method. For 
energies from 150 to 800~MeV, the $d+Be$ break-up reaction is used at Saturne to produce quasi 
monoenergetic neutrons. The deuteron beam intensity is measured by activation of a carbon foil~\cite{Que} 
and the neutron flux is deduced from $d+Be$ break-up cross-sections~\cite{Lec}.

The results are displayed in Fig.~\ref{fig:figeffi1} and Fig.~\ref{fig:figeffi2}. For the Saturne 
measurement, the neutron flux attenuation along the 7 to 8.5~meters of flight path in air and in the veto 
scintillator is the same during the efficiency measurements and the real experiment: therefore it has not 
to be corrected for. On the contrary, experimental points measured at Bruy\`eres and Uppsala have to be 
corrected to take into account the difference in neutron flux attenuation due to the difference in 
distance and the absence of the veto detector. The corrected efficiencies used in the data analysis are 
represented by the solid line in Fig.~\ref{fig:figeffi1} and Fig.~\ref{fig:figeffi2} for DEMON and DENSE 
detectors respectively. The efficiency calculations performed with O5S~\cite{O5S} (diamonds) and a 
modified version of the KSU (triangles) Monte-Carlo codes, agree fully with experimental results. In the 
original KSU code~\cite{KSU} some reactions are not taken into account and at 90~MeV the sum of the cross 
sections over all the inelastic processes gives 165~$mb$, that is 90~$mb$ less than the global cross 
section measured by Kellog~\cite{Kellog}. In our modified KSU code, the total inelastic cross-section has 
been normalised to the Kellog measurements by an appropriate increase of the $(n,\alpha)$ light response. 
This is justified by the fact that the missing reactions in the KSU model produce essentially recoil 
nuclei like deuterons, tritons, alpha, lithium or boron whose light production is close to that of alphas. 
The line presented in Fig.~\ref{fig:figeffi1} is the efficiency used for all DEMON detectors. The DENSE 
detectors being used only between 2 and 14~MeV, efficiency has been determined by measurements only in the 
2-17~MeV range (see Fig.~\ref{fig:figeffi2}). The efficiency was assumed to be the same for all the 
detectors of the same type.

The beam is monitored by the start scintillator located in front of the target. Uncertainties on the 
cross-section determination are due to statistical and systematic errors. Systematic errors come mainly 
from the subtraction of accidental coincidences, gamma rejection and efficiency determination. The error 
on the latter mainly depends on the absolute calibration procedure used at the different accelerators and 
are estimated to be 10\% for the Bruyeres measurement (knowledge of the standard detector), 4\% at Uppsala 
($n-p$ cross-section) and 10\% at Saturne ($d+Be$ cross-section). The values are summarized in table 
~\ref{table:erreurBE}. 

\begin{table}[htb]
	\centering
	\begin{tabular}{|c||c|c|c|} 
		{\bf {Error origin }} & {\bf{2-20 MeV}} & {\bf{20-100 MeV}} & {\bf{100-400 MeV}} \\ 
\hline\hline
		subtraction of accidental coincidences & $5.8\%$ & $5.8\%$ & $5.8\%$\\
		gamma rejection & $2.9\%$ & $2.9\%$ & $2.9\%$\\
		efficiency determination & $10\%$ & $4\%$ & $10\%$\\ \hline\hline
		TOTAL uncertainty & $11.9\%$ & $7.6\%$ & $11.9\%$ \\ 
	\end{tabular}
	\caption{Estimations of systematic errors in the time-of-flight method.}
	\label{table:erreurBE}
\end{table}

The neutron energy resolution depends on a time and a geometrical component and is given by :
\begin{equation}
	\frac{\sigma_{E}}{E} = \gamma \left( \gamma + 1 \right) {\left[ {\left( \frac{\sigma_{l}}{l} 
\right)}^{2}
	+ {\left( \frac{\sigma_{t}}{t} \right)}^{2} \right]}^{1/2}
	\label{eq:equationresolution}
\end{equation}
with $\gamma$ the Lorentz factor\\
      $\frac{\sigma_{t}}{t}$ the time resolution\\
      $\frac{\sigma_{l}}{l}$ the geometrical component\\
The time resolution ($\sigma_{t}=1.5~ns$ and 2.0 ns for DEMON and DENSE detectors respectively) is 
estimated by the measurement of the FWHM of the prompt gamma peak on the time-of-flight spectra. The 
geometrical component comes from the target thickness (1 to 3 cm) and from the size of the sensitive area 
of the detector. 
The interaction probability being constant as a function of the depth the standard uncertainty is 
$\sigma_{l}=\frac{L}{2\sqrt{3}}$~\cite{ISO}. Thus $\sigma_{l}=6~cm$ and 1.5~cm for DEMON and DENSE 
detectors respectively.   

\begin{figure}[htb]
	\vspace{-0.4cm}
	\includegraphics[width=8.0cm]{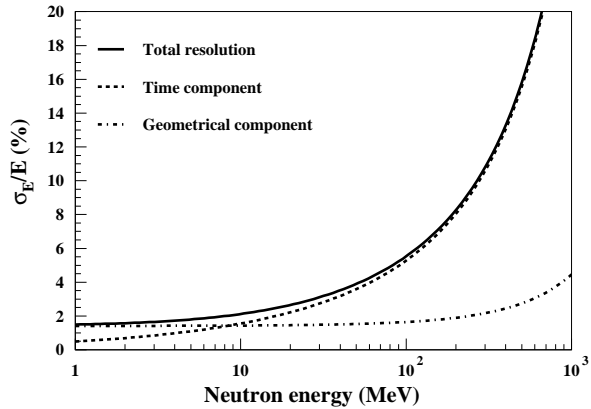}
	\caption{Geometrical and time contributions to the energy resolution as a function of neutron 
energy for a DEMON detector. The time-of-flight length is 8.5~m, the time uncertainty 1.5~ns and the 
length uncertainty is 6~cm.}
	\label{fig:figureresolution}
\end{figure}

The energy resolution is plotted as a function of the energy in figure~\ref{fig:figureresolution}.
It appears clearly that above 400~MeV, this time-of-flight method doesn't allow neutron energy 
measurement with a resolution better than $12\%$. Therefore another complementary detection system has 
been developed and is described in the next section.

\subsection{Proton recoil spectrometer.}

\begin{figure}[htb]
	\includegraphics[width=8.cm,angle=-90.]{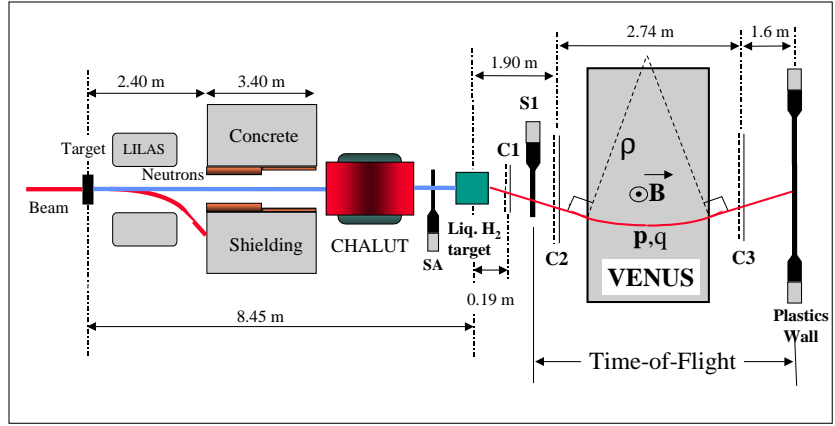} 
   	\caption {Principle of the high energy neutron measurement ($E>200$~MeV) with the spectrometer, 
shown here at $0^{\circ}$.}
   	\label{fig:dispospect}
\end{figure}

High energy (i.e. above 200 MeV) neutrons are detected using $(n,p)$ scattering on a liquid hydrogen 
converter and detecting the emitted proton in a magnetic spectrometer (Fig ~\ref{fig:dispospect}). We 
present here only a comprehensive description of the measurement with a special emphasis on the 
modifications compared to the the first experiment at $0^{\circ}$ which is detailed in ref~\cite{Mar}.   

The spectrometer is composed of the dipole magnet VENUS, which generates a 0.4~T field, and of 3 multiwire 
proportional chambers, C$_{1}$, C$_{2}$ and C$_{3}$, of respective active area 20x20cm$^2$, 80x40cm$^2$ 
and 100x80cm$^2$. Each chamber is composed of 2 sets of wires equipped with PCOS2 electronics allowing the 
localization in the $X-Y$ plane. The wires are spaced by 1.27mm for C$_{1}$ and 2mm for C$_{2}$ and 
C$_{3}$. 

The acquisition is triggered by the coincidence between the plastic scintillator S1 and the large wall of 
NE102 plastic scintillators behind VENUS. This wall is made of 20 horizontal slats with a photomultiplier 
on each side.

A second dipole magnet, CHALUT, deviates in the horizontal plane the charged particles created in the target, 
in the concrete or in air. The field integral of this magnet was $0.4 Tm$ during our experiment. The thin 
plastic scintillator SA tags the possible charged particles remaining in front of the liquid hydrogen target. 
The spectrometer (hydrogen target and detectors) and CHALUT are placed on platforms which could rotate from 0 
to $85^\circ$ only since, for larger angles, very few neutrons with energies higher than 400~MeV are expected. 
A single setting of the magnetic field in VENUS ($0.4 T$) was chosen during the standard measurements. The 
center of the chamber C$_{3}$ was shifted by $40~cm$ to the left with respect to the beam axis in order to 
optimize the detection of deflected protons.
    
The trajectory reconstruction of the charged particles emitted by the hydrogen target is deduced from the 
impact coordinates in C$_{1}$, C$_{2}$ and C$_{3}$. The well known magnetic field inside VENUS gives their 
momentum. The geometric calibration of the multiwire chambers was done using a low intensity 
$800~MeV$ proton beam without the hydrogen target and successive magnetic fields of $-0.2$, $0.$, $0.2$ 
and $0.4$ Tesla.

The liquid hydrogen target is a cylinder of 12.8~cm in diameter. It has a useful thickness of 
$0.94~g/cm^{2}$ and is located 8.45 meters from the production target, covering an angular aperture of 
$0.43^{\circ}$. The entrance and outgoing windows are titanium foils, 100~$\mu m$ thick .

\begin{figure}[htb]
	\includegraphics[width=8.cm,angle=-90.]{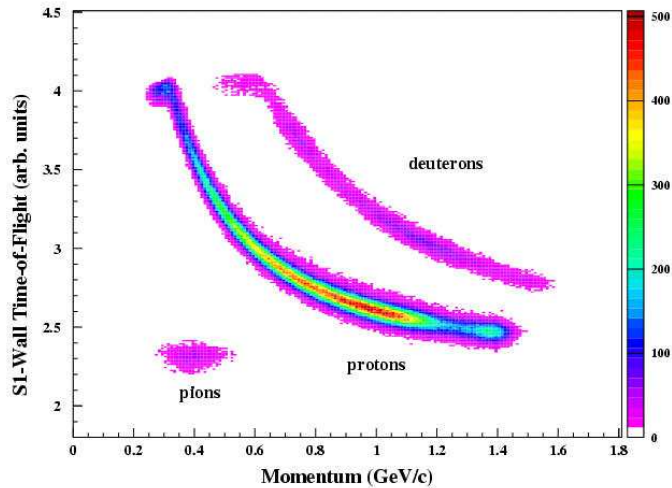}
   	\caption{Bidimensional spectra to identify protons, deuterons and pions produced in the hydrogen 
target.}
   	\label{fig:tofimpul}
\end{figure}

Several types of charged particles are created in the H$_{2}$ target through the following processes : 
np$\rightarrow$np, np$\rightarrow$np$\pi^{o}$, np$\rightarrow$pp$\pi^{-}$, np$\rightarrow$d$\pi^{o}$, 
np$\rightarrow$2n$\pi^{+}$. The charged particles with different masses are identified using the 
biparametric representation of the S1-Wall time-of-flight versus the momentum measured with the wire 
chambers and VENUS (Fig.~\ref{fig:tofimpul}). 

\DeclareGraphicsRule{.jpeg}{eps}{.jpeg.bb}{`convert #1 'eps:-' }

The incident proton beam is monitored by two telescopes viewing a 50 $\mu m$ mylar foil placed upstream in the beam ($\sim 20 m$ from the target). The absolute calibration of these telescopes is obtained by a 
comparison with the activation of a carbon sample~\cite{Que} measured in a dedicated run. A calibration 
with activation of Al foils was also done and gives a very comparable result. 
 
The response function of the spectrometer which takes into account the contribution of elastic and 
inelastic processes arising in the hydrogen target has been measured with quasi-monoenergetic neutrons 
produced by the break-up of deuterons or $^3$He beams on a Be target. From various beam energies, neutrons between 0.2 and 1.6 GeV were produced for this calibration. The neutron flux is obtained from the known (n,p) elastic scattering cross-sections~\cite{CEX} and the normalized response functions are then used to unfold the measured proton spectra. This procedure gives the normalized neutron energy distribution. It is described in detail in ref~\cite{Mar}. The maximum energy available at Saturne (1.6~$GeV/A$ for $^{3}He$) and the growing importance of the inelastic processes set a limit to this unfolding procedure. 

\begin{center}
   \begin{table}[htb]
      \centering
	\begin{tabular}{|c||c|c|} 
		 		      & \multicolumn{2}{c|}{\bf{Beam Energy}}\\ \cline{2-3}
		{\bf {Error origin }} & {\bf{0.8 and 1.2~GeV}} & {\bf{1.6~GeV}}\\ \hline\hline
		Beam monitoring & $\leq$ 5.8$\%$& $\leq$ 5.8$\%$ \\
		Spectrometer response function & $\leq$ 4$\%$& $\leq$ 11.5$\%$ \\
		Unfolding procedure & $\leq$ 5.8$\%$& $\leq$ 8.6$\%$ \\ \hline\hline
		TOTAL uncertainty& $\leq$ 9.1$\%$ & $\leq$ 15.5$\%$ \\ 
	\end{tabular}
	\caption{Estimations of systematic errors as a function of incident energy.}
	\label{table:erreurHE}
   \end{table}
\end{center}

Systematic errors in this method arise mainly from the beam calibration, the spectrometer response function and the unfolding procedure. The estimations of these 3 errors are given in table~\ref{table:erreurHE}. They are less than $10\%$ at 800 and 1200~MeV but reach $15.5\%$ at 1600~MeV due to the increasing inelastic contribution. Error bars on the results presented in this paper take into account only statistical uncertainties except at the very low energies where the increase of uncertainty associated with the proximity of the detection threshold is added.

\subsection{Targets}

The same targets were used for both the time-of-flight and spectrometer methods. Because of the low beam 
intensity imposed by the detection of the incident proton with the SC scintillator, we had to use rather 
thick targets in order to keep a significant counting rate. They were 3cm diameter cylinders made of natural material with thickness in the centimeter range, shown in Table~\ref{table:ep}, depending on the elements.
 
\begin{center}
\begin{table}[htb]
   \centering
	\begin{tabular}{||c||c|c|c|c|c|c||}
	\hline\hline 
	{\bf {Target}} & Al & Fe & Zr & W & Pb & Th \\ \hline
	{\bf {Thickness (cm)}} 	& {\bf {3}} & {\bf {3}} & {\bf {3}} & {\bf {1}} & {\bf {2}} & {\bf {2}}\\ 
	\hline\hline
	\end{tabular}
	\caption{Target thicknesses for the different materials used in this experiment.}
	\label{table:ep}
   \end{table}
\end{center}

\section{Double-differential cross-sections}

\subsection{Experimental results}

\begin{figure}
	\begin{minipage}[c]{.48\linewidth}
	\begin{center}
	\includegraphics[width=5.8cm]{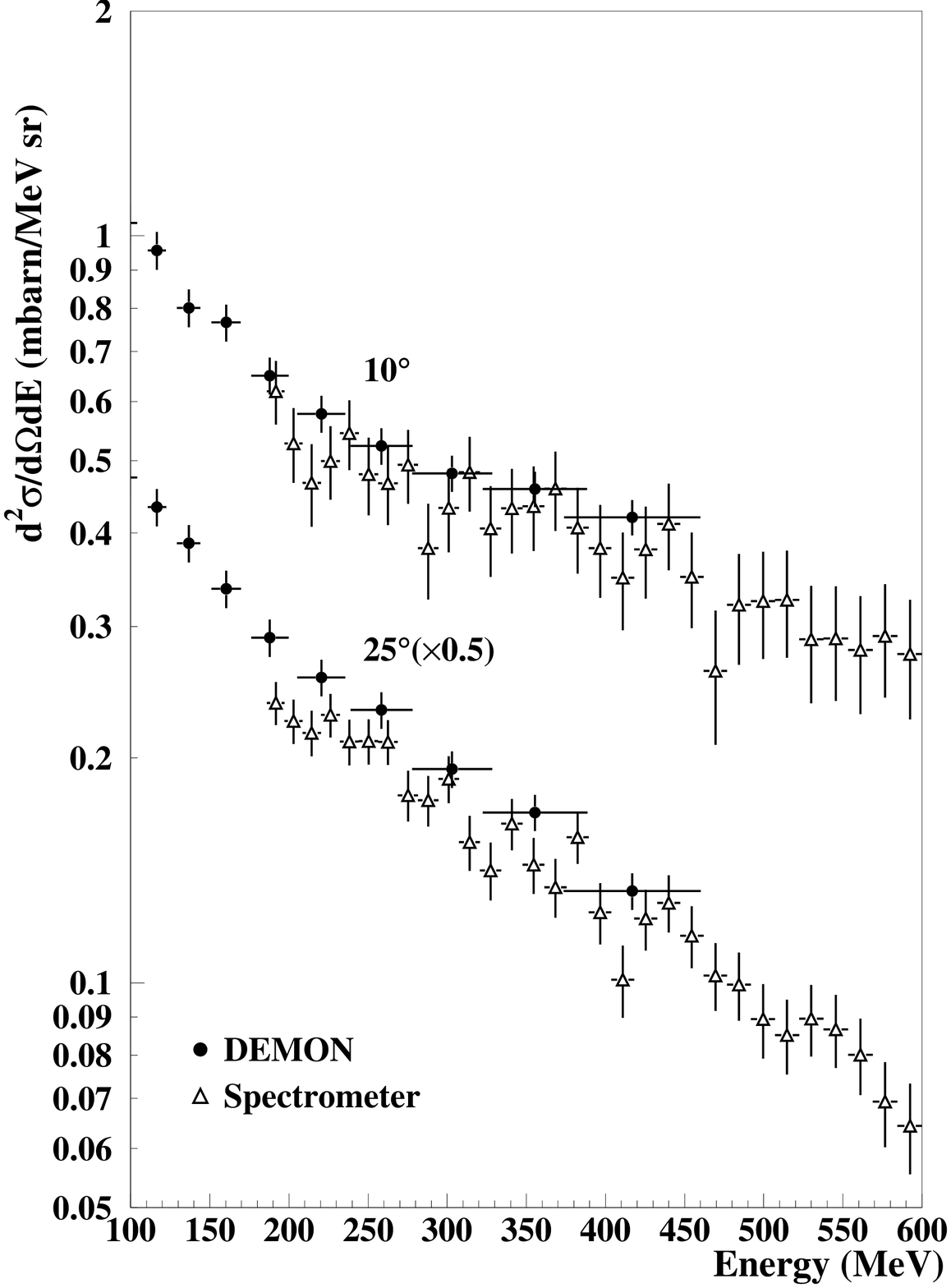}
	\end{center}
	\caption{Results obtained at 1600 MeV on the iron target with the DEMON detectors and the 
spectrometer in the overlapping region at 10 and 25$^{\circ}$.}
	\label{fig:Fereche}
	\end{minipage}
	\hfill
	\begin{minipage}[c]{.48\linewidth}
	\begin{center}
	\includegraphics[width=5.8cm]{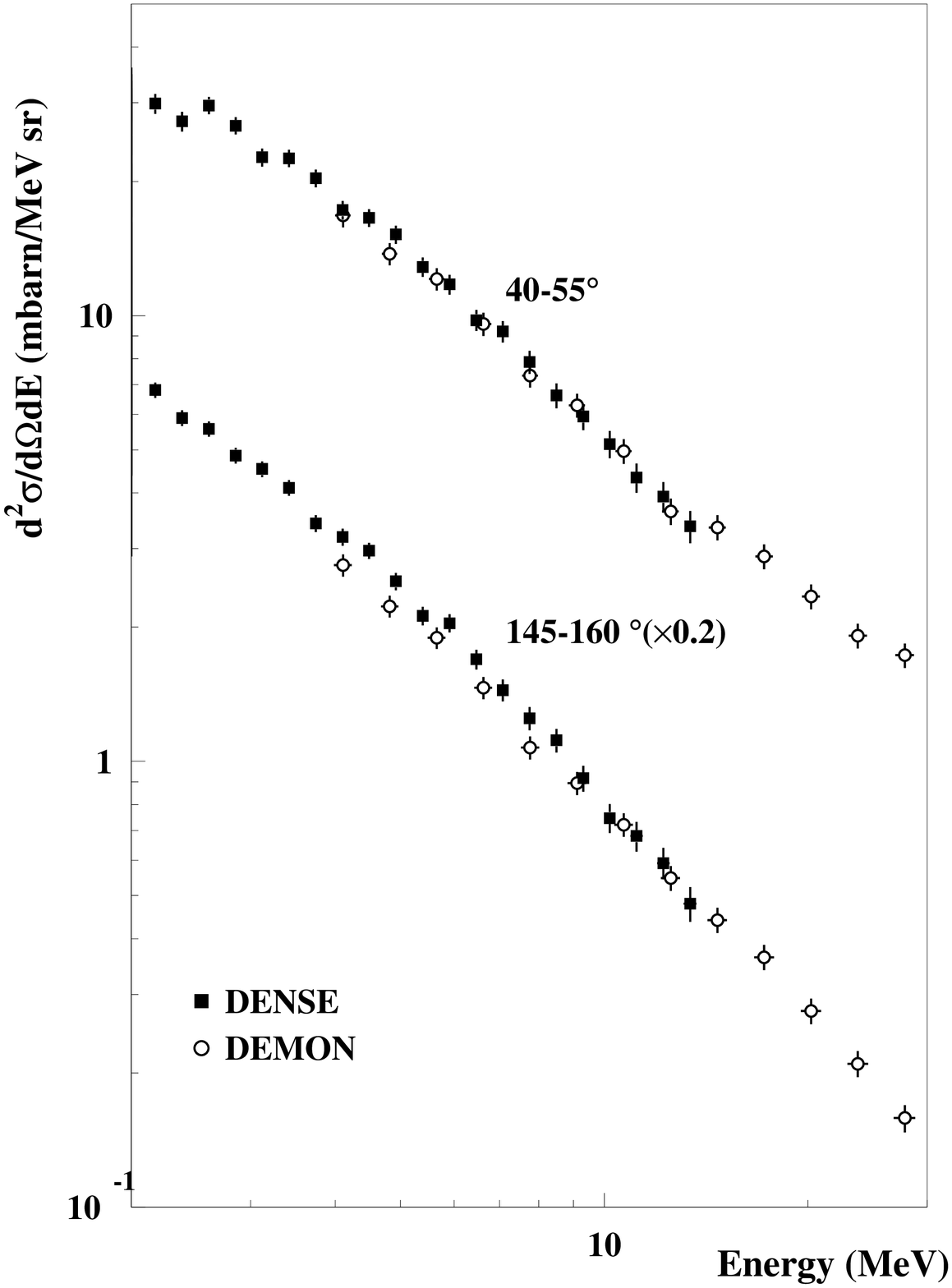}
	\end{center}
	\caption{Id. but for the overlapping region of the DENSE and DEMON detectors at close angles 40-55$^{\circ}$ and 145-160$^{\circ}$ since evaporation neutrons are emitted nearly isotropically.}
	\label{fig:Ferecle}
	\end{minipage}
\end{figure}

In order to show the consistency of the three different sets of detectors, figs.~\ref{fig:Fereche} and \ref{fig:Ferecle} display details of the double-differential cross-sections obtained for a Fe target at 1600 MeV. In fig.~\ref{fig:Fereche} data obtained at 10 and 25$^{\circ}$ with the DEMON detectors (filled circles) and the spectrometer (squares) are shown. It can be seen that in the overlap regions, i.e.between 200 and 400 MeV, the data are compatible within the error bars. Actually, 1600 MeV corresponds to the worst case since, as mentioned above, the spectrometer unfolding procedure is approaching its limits of reliability. Other examples of the good agreement between both methods regarding Pb at 800 and 1200 MeV were shown in~\cite{XL}.  In fact, for all the measurements, the data from the DEMON detector and spectrometer always agree within less than 15$\%$ at 1600 MeV and 10$\%$ at lower energies. Concerning the comparison of spectra obtained between 4 and 12 MeV with the DENSE and DEMON detectors, measurements at the same angles were done only for Pb at 1200 MeV and also shown in \cite{XL}. However, since below 15 MeV neutrons mostly come from an evaporation process which is practically isotropic in the laboratory system, results obtained at near angles can be compared. This has been done in fig.~\ref{fig:Ferecle} where 25 and 
145$^{\circ}$ DEMON spectra are plotted together with respectively 40 and 160$^{\circ}$ DENSE ones. Both sets of data are consistent and agree within less than 10\%. This appears to be verified whatever the target and the energy. Therefore, a single set of data merging the different measurements by taking their mean values in the overlap regions has been processed and will be shown in the following. 

\begin{figure}
	\hspace*{-1.2cm}
	\begin{minipage}[c]{17.5cm}
	\begin{center}
	\includegraphics[width=5.7cm]{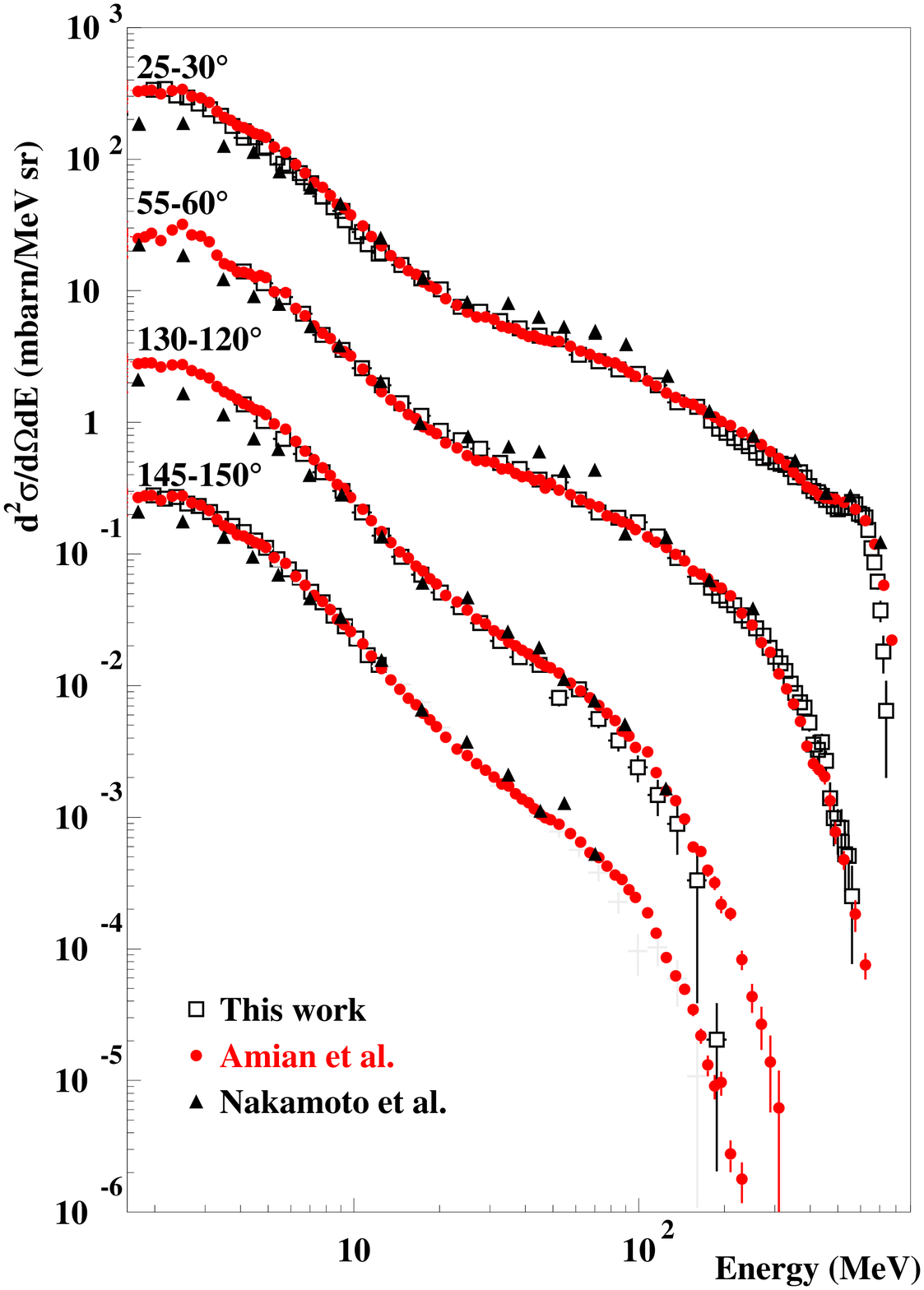}
	\includegraphics[width=5.7cm]{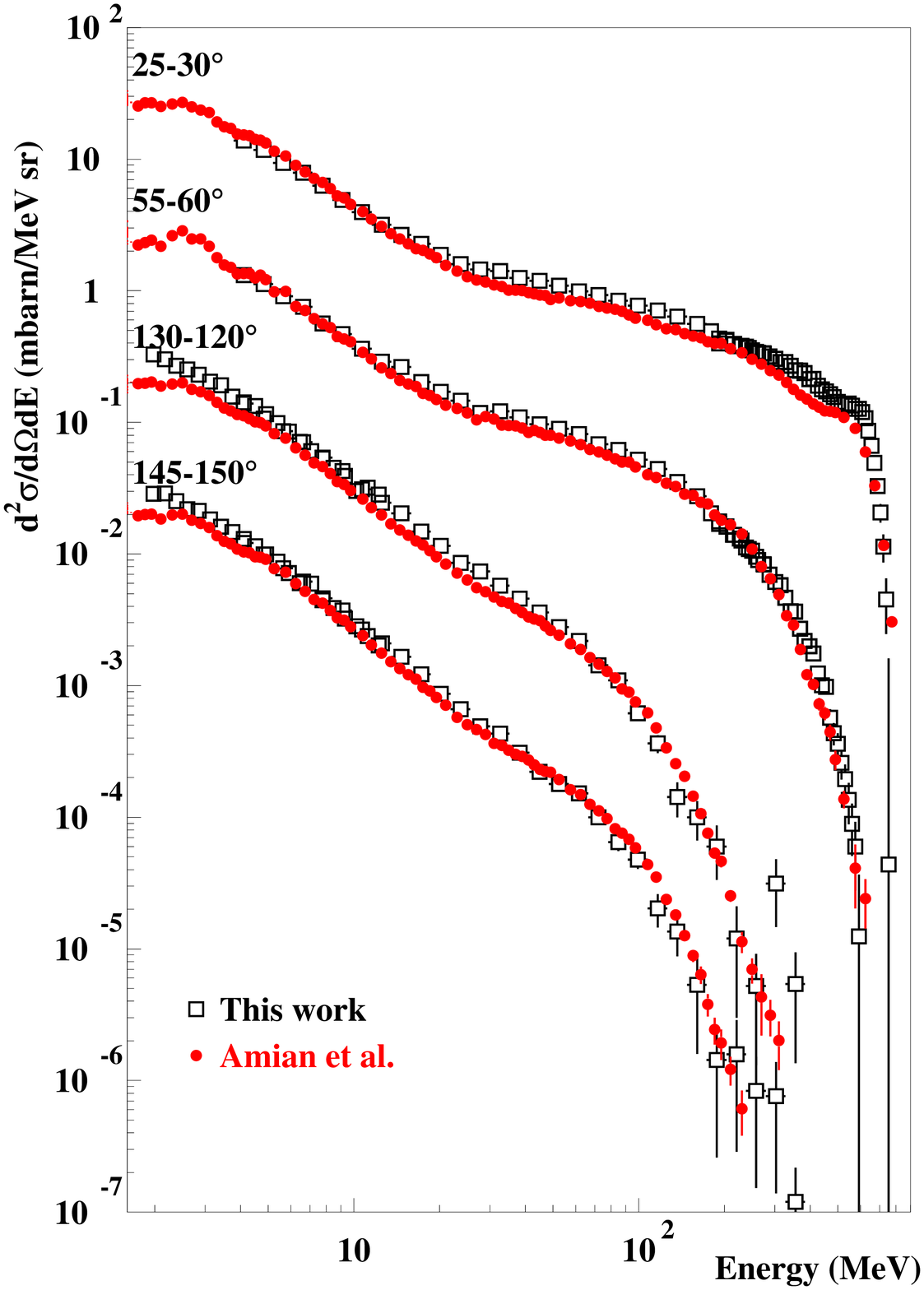}
	\includegraphics[width=5.7cm]{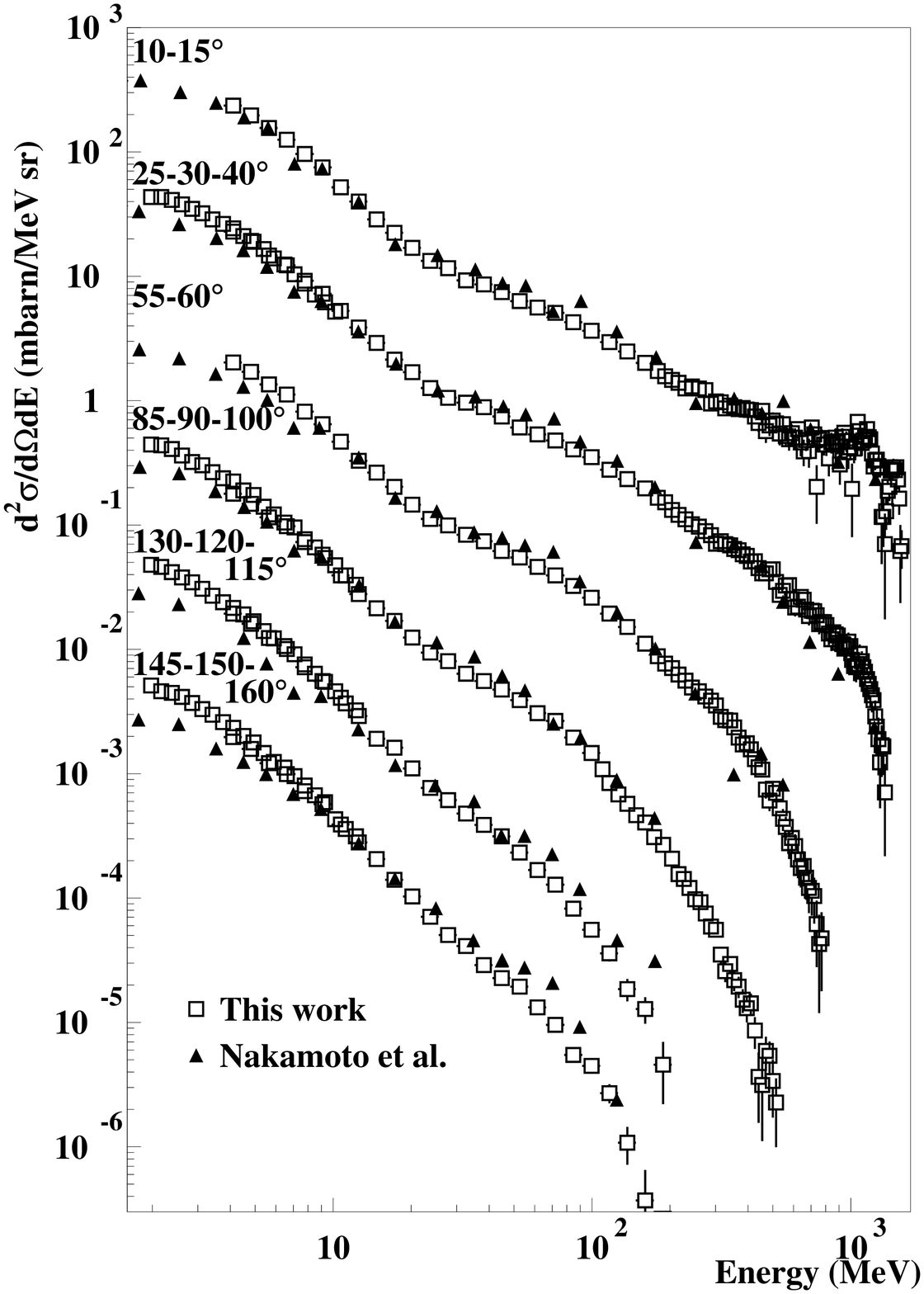}
	\end{center}
	\caption{{\small Comparison of our results with the data from Amian et al.~\protect\cite{Ami} and Nakamoto et al.~\protect\cite{Naka} at different close angles; left : for Pb at 800 MeV; center : for Fe at 800 MeV; right : for Pb at 1600 MeV. Each successive curve, starting from the smallest angle, is scaled by a multiplicative factor of 10$^{-1}$.}}
	\label{fig:CompAmiNak}
	\end{minipage}
\end{figure}

\begin{figure}
	\begin{center}
	\begin{minipage}[c]{15.6cm}
	\begin{center}
	\includegraphics[width=5.6cm]{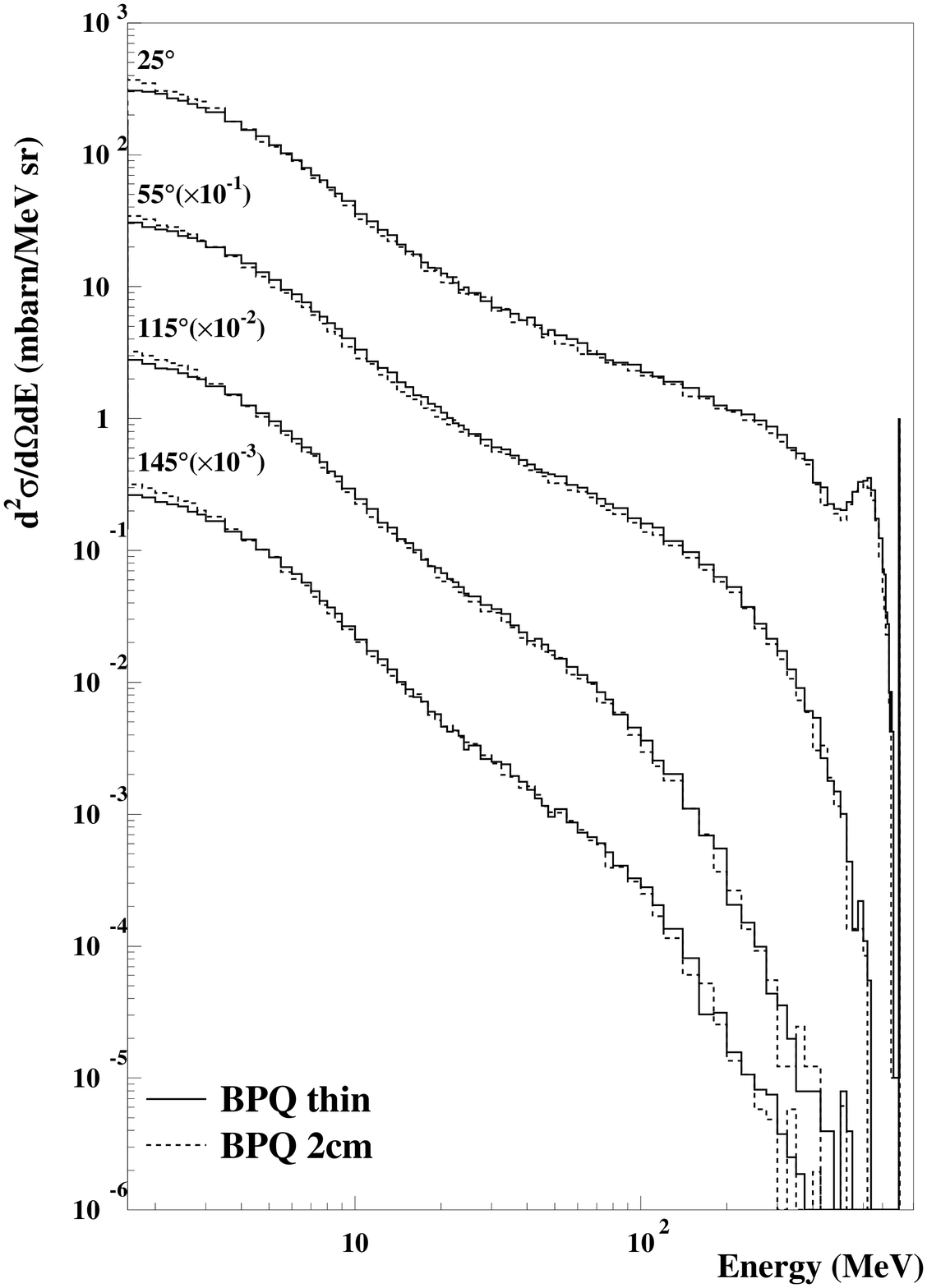}
	\includegraphics[width=5.6cm]{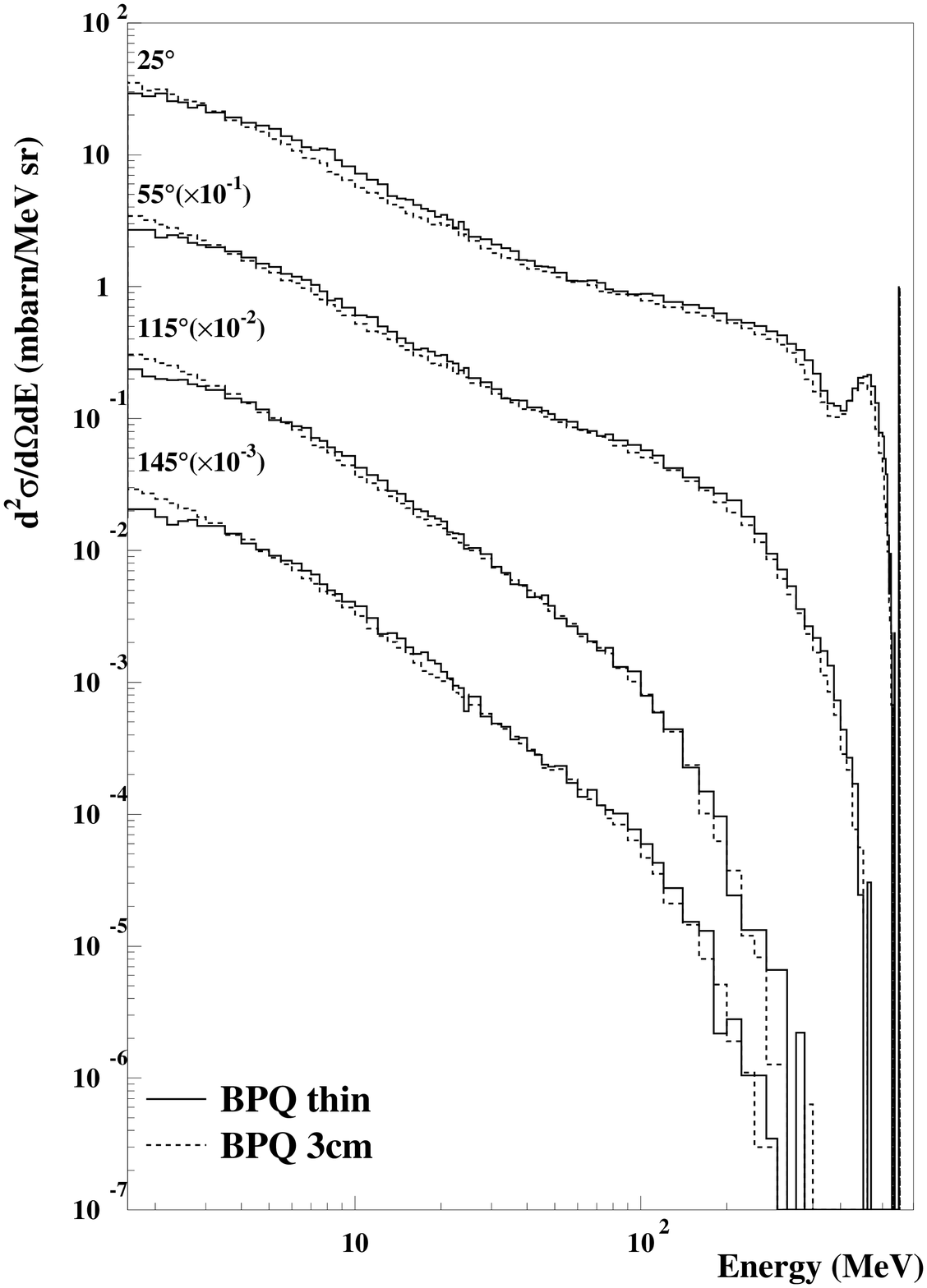}
	\end{center}
	\caption{Effect of the target thickness at 800 MeV for Pb (left) and Fe (right) : calculations done with the LAHET code~\protect\cite{LAHET} for an infinitely thin target (solid line) and a 2cm Pb and 3cm Fe target respectively (dashed line).}
	\label{fig:pbfe_ep}
	\end{minipage}
	\end{center}
\end{figure}

Fig.~\ref{fig:CompAmiNak} shows comparisons of our data with previously obtained ones by Amian et 
al.~\cite{Ami} and Nakamoto et al.~\cite{Naka} using time-of-flight. In each case we compare the data at the closest possible angles, adding when appropriate results from DEMON and DENSE detectors at two different angles. Actually, our 10, 25-40, 55, 85-100, 115-130 and 145-160$^{\circ}$ data are displayed together with the previous ones at respectively 15, 30, 60, 90, 120 and 150$^{\circ}$. Our Pb measurements at 800 MeV (left) fully agree with Amian ones, as already noticed in~\cite{XL}, while Nakamoto cross-sections are systematically lower at low neutron energy and higher at energies between 10 and 100 MeV. For Fe (center), only data from~\cite{Ami} are available and we observe a slightly less good agreement between the two works. However, it should be stressed that contrary to \cite{Ami} our (and ref~\cite{Naka}) targets are not really thin and secondary reactions increase the number of low energy neutrons (as discussed below). This effect is visible only below 4 MeV and appears to be larger for a 3cm thick Fe target than for a 2cm thick Pb one, as shown by the simulations in fig.~\ref{fig:pbfe_ep}. This could explain why we measure more neutrons than Amian et al. at low energies in the case of iron. At 1600 MeV for Pb (fig.~\ref{fig:CompAmiNak} right), we have compared our data to ref~\cite{Naka} data obtained at 1500 MeV. This is possible since, from our 1200 and 1600 MeV measurements, we could infer that cross-sections should differ by less than 5\% between 1500 and 1600 MeV, apart from the high energy part of the spectra at very forward angles. As observed at 800 MeV, we get higher cross-sections below 7 MeV and lower ones at intermediate energies. Since the thickness of the targets are the same in both cases, this can be understood only by differences in the neutron detector efficiency determination. As mentioned earlier, at intermediate energies, our experimentally determined efficiency~\cite{Thun} is higher than the one calculated using the standard KSU code and thus we used a modified version. As far as we know, it was the standard version of KSU that was used to determine the detector efficiency in~\cite{Naka}. This could explain the discrepancy. At high energies, with the spectrometer, we obtain a much better energy resolution than in ref.~\cite{Naka} (because of their limited flight path) that allows us to distinguish structures due to direct reactions at forward angles.

As already mentioned, the thickness of our targets induces some distortion in the double-differential spectra because of the slowing down of the incident proton and the probability that some of the energetic emitted particles may undergo secondary reactions. The first point was discussed in~\cite{XL} and results in a slight shift and a broadening of the quasi-elastic and inelastic peaks at very forward angles. The second effect is expected to lead to a depopulation of the high and intermediate energy parts of the spectra and an increase of the number of low energy emitted neutrons. Calculations using the LAHET high-energy transport code system~\cite{LAHET} (using Bertini as intra-nuclear cascade model and pre-equilibrium) were performed for both a target with the actual geometry and an infinitely thin one in order to assess the order of magnitude of the effect. In fig.~\ref{fig:pbfe_ep}, results are shown for Pb and Fe at 800 MeV. It can be seen that the difference is very small for the 2cm lead target, for which it is significant only between 2 and 3 MeV, and a little larger for the 3cm iron one. In the latter case, the disappearance of intermediate energy neutrons is also perceptible. Similar results are found with the other targets and at other energies, the effect being maximum for Fe and W.

All the measured angular distributions are presented together with the model calculations in the next section. Data have been taken at 0.8, 1.2 and 1.6~GeV on Pb and Fe targets and at 1.2~GeV on Th, Pb, W, Zr, Fe and Al targets at 0, 10, 25, 55, 85, 130 and 160$^{\circ}$.
   
\subsection{Comparison with models}

Spallation reactions are generally described by a two step mechanism: a first stage in which successive hard collisions between the incident particle and the individual nucleons of the target nucleus lead to the emission of a few fast nucleons, then, the decay of the excited remnant nucleus by emission of low energy particles or, sometimes for heavy nuclei, by fission. The first step is generally described by Intra-Nuclear Cascade models while evaporation-fission models are used for the second one. Some authors introduce a pre-equilibrium stage between intra-nuclear cascade and de-excitation. In high energy transport codes, the most widely used intra-nuclear cascade model is the old Bertini~\cite{Ber} one dating from 1963. However, several other models are available, such as the Isabel~\cite{Isa} and the Cugnon~\cite{Cugn,Cugn1} INCL models, which have brought some improvements in the physics. The most widely used evaporation model in the domain of spallation reactions is the Dresner model~\cite{Dres}, usually associated with the Atchison~\cite{Atch} fission model.

The high energy part of the neutron spectra enables one to directly probe the intra-nuclear cascade models. Low energy neutrons, which are the majority of the neutrons produced in spallation reactions, are emitted during the evaporation process. However, their number mainly depends upon the intra-nuclear cascade stage since the cascade determines the initial excitation energy of the decaying hot residue and, therefore, the number of evaporated particles. Actually, evaporation neutron spectra for a given excitation energy are not expected to depend very much on the evaporation-fission model contrary to light charged particle spectra or to residual nuclei production for which emission barriers and competition between the different decay modes are not so well established. This is why we have made comparisons with different intra-nuclear cascade models in order to test their validity and understand their differences, using always the same evaporation-fission model. 

All the calculations discussed in the following have been done with high energy transport codes in which the actual thickness and diameter have been taken into account. In order to have sufficient statistics, calculations were done for angular bins of 5 degrees (except at 0 degrees where it is only 2.5 degrees). The INCL model does not predict a correct total reaction cross-section mainly because the diffuseness of the nuclear surface is not taken into account. Therefore, the INCL calculations were renormalized to the total reaction cross-sections given by the Bertini model which appear to be in very good agreement with experimental values from ~\cite{Bar}. 

\begin{figure}
	\begin{minipage}[t]{.48\linewidth}
	\begin{center}
	\includegraphics[width=5.6cm]{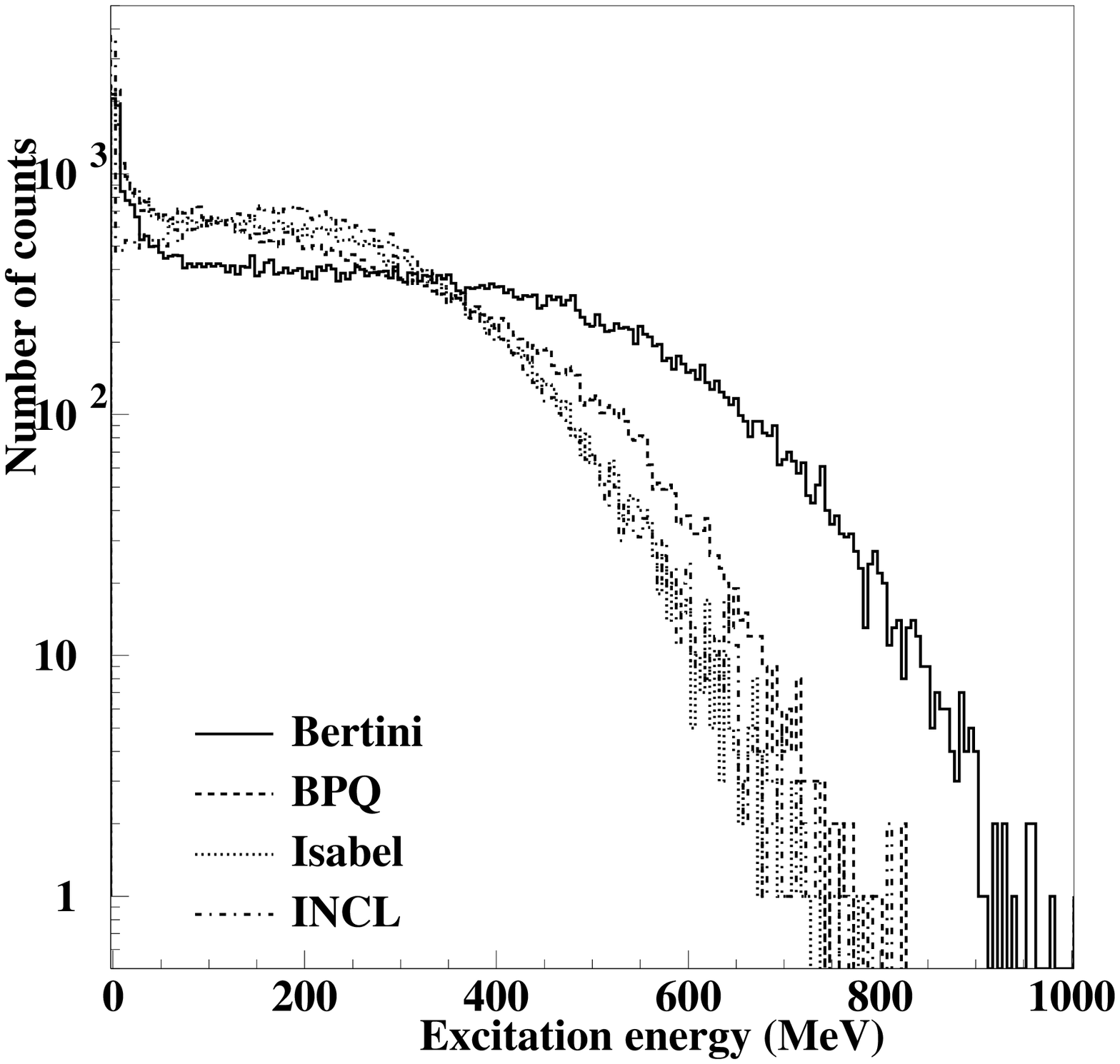}
	\end{center}
	\caption{{\small Excitation energy distribution in the p (1 GeV) + Pb reaction found with the 
Bertini ({\it solid line}), Bertini+pre-equilibrium ({\it dashed line}), Isabel ({\it dotted line}) or 
INCL ({\it dashed-dotted line}) intra-nuclear cascade models.}}
	\label{fig:pbex}
	\end{minipage}
	\hfill
	\begin{minipage}[t]{.48\linewidth}
	\begin{center}
	\includegraphics[width=5.6cm]{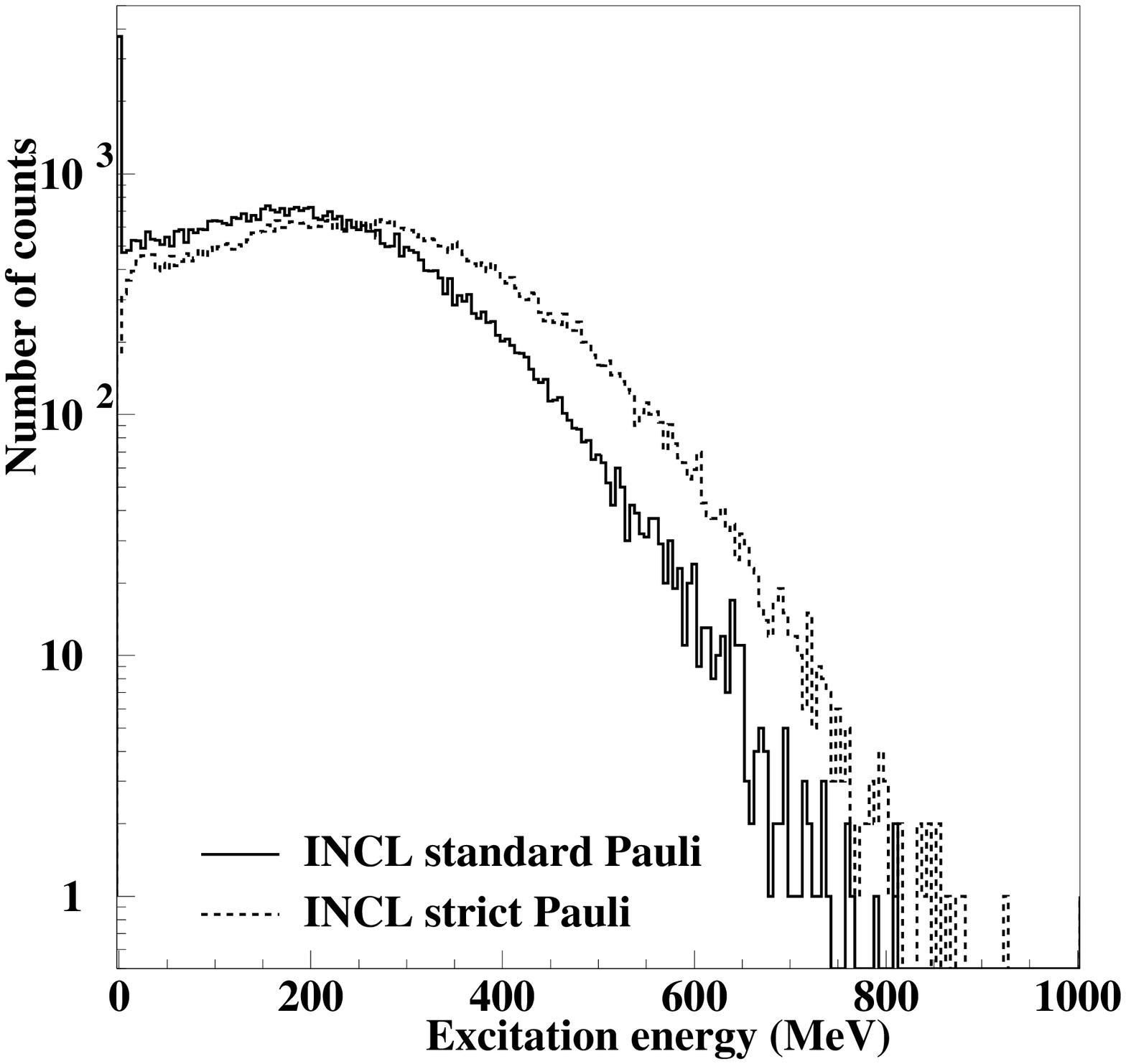}
	\end{center}
	\caption{Excitation energy distribution in the p (1 GeV) + Pb reaction calculated with the Cugnon INCL model using the standard ({\it solid line}) or a strict Pauli blocking ({\it dashed line}).}
	\label{fig:pauli}
	\end{minipage}
\end{figure}

In~\cite{XL}, for lead, we presented calculations performed with the TIERCE~\cite{TIERCE} high-energy transport code system developed at Bruy\`eres-le-Ch\^atel (which is very similar to LAHET) using either the Bertini or the Cugnon INCL model with the same evaporation-fission model (based on the Dresner-Atchison model). It was shown that, at the three measured energies, the Bertini model was largely overpredicting the experimental data while INCL was giving a rather good agreement. This was ascribed to the higher excitation energy, $E^*$, obtained at the end of the cascade stage with the Bertini calculation than with INCL. This assumption can be verified in fig.~\ref{fig:pbex} where the $E^*$ distribution obtained with Bertini (solid line) is shown to extend to much higher values than INCL (dashed-dotted line) and gives also a higher average value (265 versus 213 MeV). These calculations were performed for thin targets at 1 GeV. The same observations were also made in \cite{NESSI} and \cite{DF} where a similar plot was shown for p+Au reactions and INCL was found to give the best agreement with the excitation energy deduced from the neutron multiplicity distributions. Several reasons can explain the difference in $E^*$ between the two models : first, INCL leads to the emission of more pions than Bertini. However, the difference in average $E^*$ due to the energy carried away by the pions is only 30 MeV. Second, as mentioned in \cite{XL}, the Pauli blocking is treated in a different way. In Bertini, only collisions of nucleons with momentum larger than the Fermi momentum are allowed while, in INCL, the actual phase space occupation rate is taken into account. This leads to a less stringent condition, therefore more cascade particles can escape and make the energy remaining in the nucleus lower. This is illustrated in fig.~\ref{fig:pauli} where the $E^*$ distributions with a strict Pauli blocking (as in Bertini) and the standard one are shown. The decrease is obvious.

\begin{figure}[htb]
	\begin{minipage}[c]{15.6cm}
	\begin{center}
	\includegraphics[width=7.3cm]{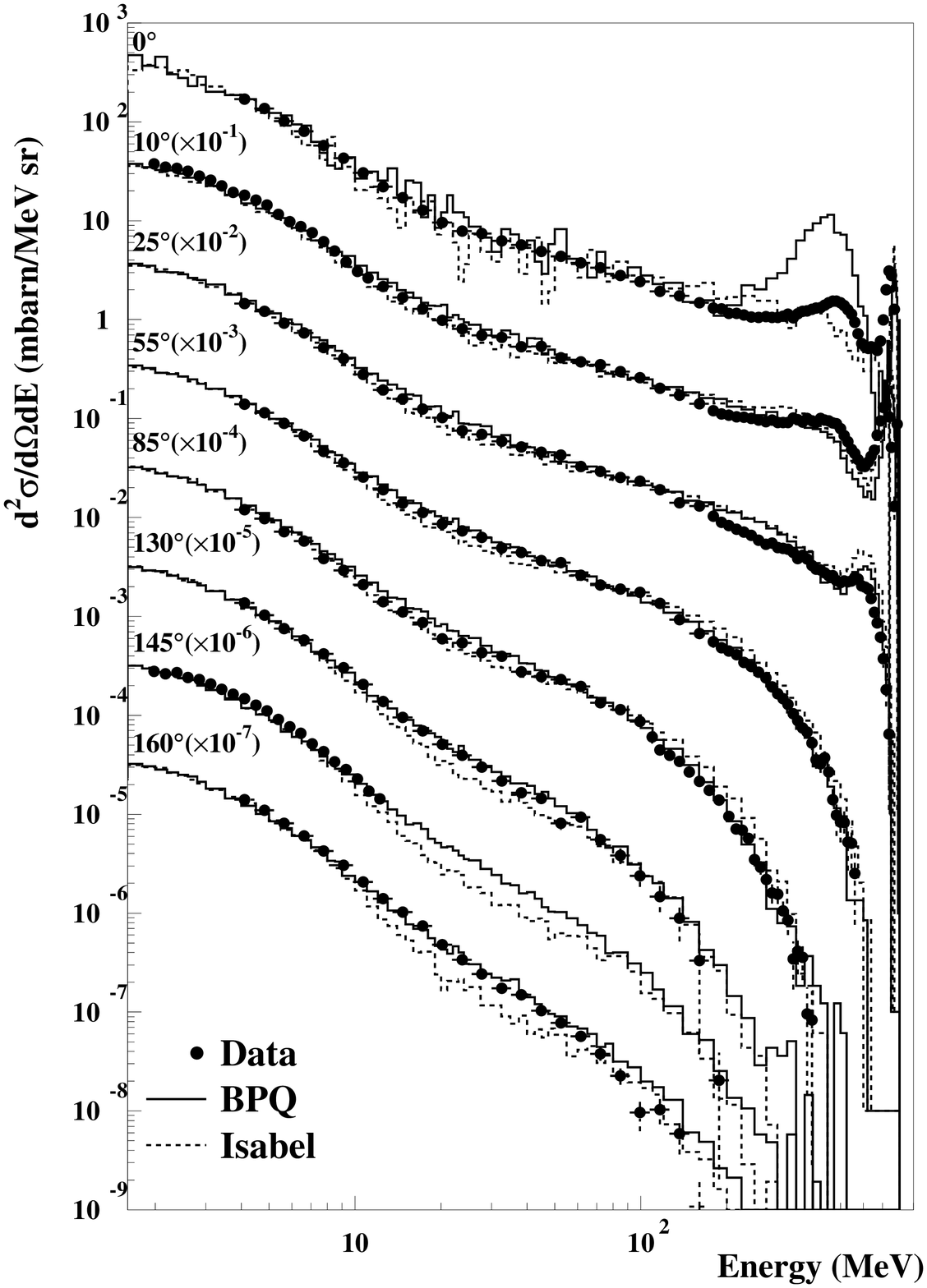}
	\includegraphics[width=7.3cm]{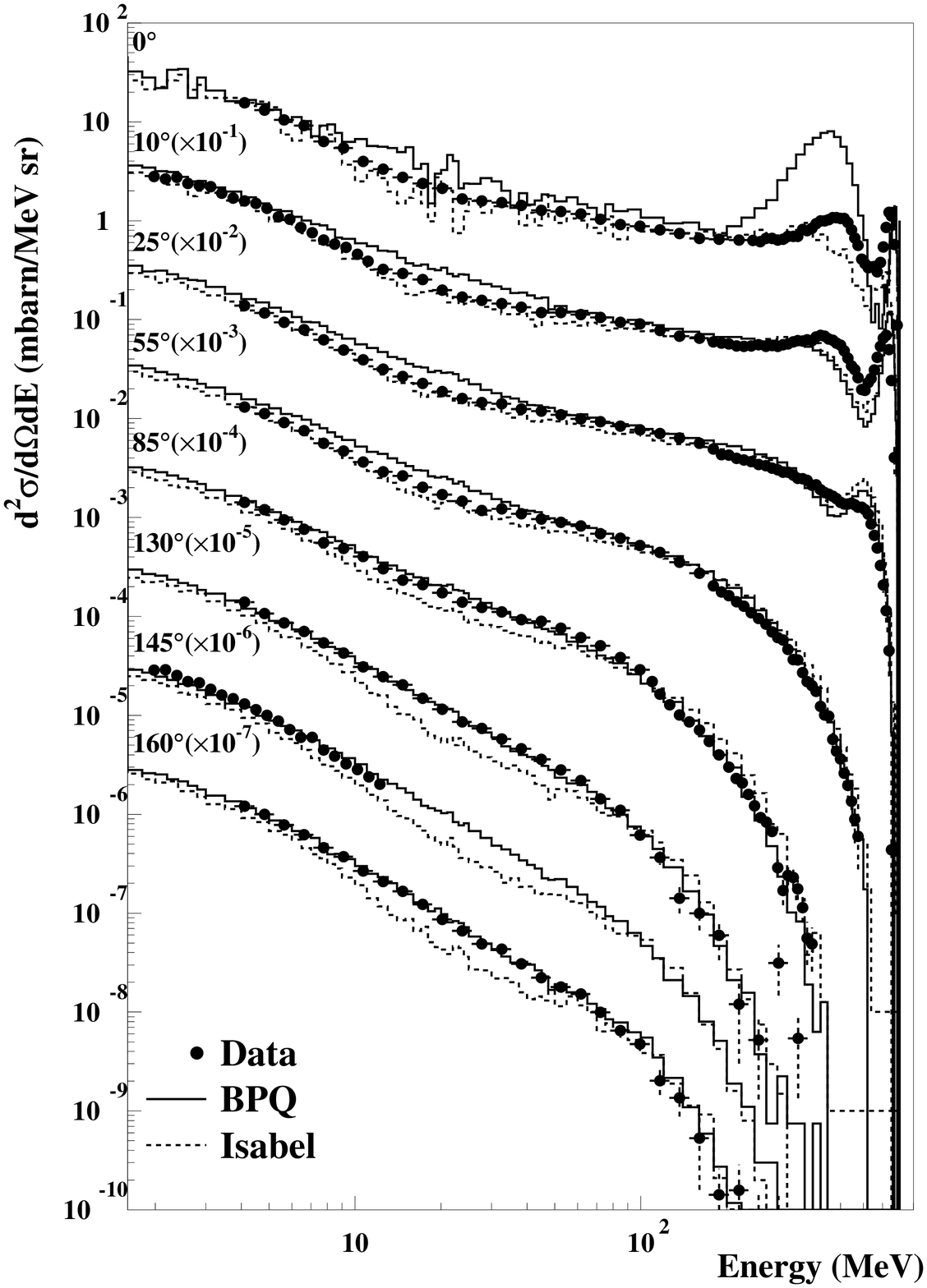}
	\end{center}
	\caption{Experimental p (800 MeV) + Pb (left) and Fe (right) neutron double-differential 
cross-sections compared with calculations performed with LAHET using either Bertini plus pre-equilibrium 
({\it solid line}) or Isabel ({\it dashed line}) intra-nuclear cascade model. Each successive curve, starting from 0$^{\circ}$, is scaled by a multiplicative factor of 10$^{-1}$.}
	\label{fig:pbfe800}
	\end{minipage}
\end{figure}

In the LAHET code system~\cite{LAHET} it is possible to add after the intra-nuclear cascade stage a 
pre-equilibrium~\cite{Preq} stage which is expected to reduce the excitation energy of the nucleus by emission of intermediate energy particles prior to the evaporation. Besides, this is also the recommended option by the LAHET authors~\cite{Prael}. Also available is the Isabel model which can be used only up to 1 GeV. As can be seen in fig.~\ref{fig:pbex}, both models lead to excitation energy distributions close to the one found with INCL. Isabel is used with the partial Pauli blocking (recommended) option, which, as in INCL, is supposed to take into account the depletion of the phase space due to the emission of cascade particles. Here, we show calculations performed with both models and the same Dresner-Atchison evaporation-fission at 800 MeV for the Pb and Fe target. In the following, Bertini plus pre-equilibrium will be referred to as BPQ. Fig.~\ref{fig:pbfe800} presents the calculated neutron spectra compared to the experimental data. It can be observed that, for Pb, the BPQ calculation reproduces very well the data, except at very forward angles and high neutron energies where the peak corresponding to the excitation of the $\Delta$ resonance appears much too high. This is a deficiency of the Bertini intra-nuclear cascade model, already pointed out in~\cite{Delta} as due to a bad parameterisation of the $N N \rightarrow N \Delta$ reaction angular distribution. The problem does not exist with Isabel. Both models correctly predict the low energy part of the spectra. This can be understood by the respective excitation energy distributions being similar in their extension to the one found with INCL (see fig.~\ref{fig:pbex}). The high energy neutrons above 85$^{\circ}$ are also well reproduced by both calculations but Isabel underestimates cross-sections at backward angles in the intermediate energy region. For iron, Isabel presents the same features as for lead while BPQ now overpredicts low and intermediate energy neutron production at forward angles, indicating that the angular distribution of pre-equilibrium neutrons is probably too much forward-peaked.

\begin{figure}[htb]
	\begin{minipage}[c]{15.6cm}
	\begin{center}
	\includegraphics[width=7.3cm]{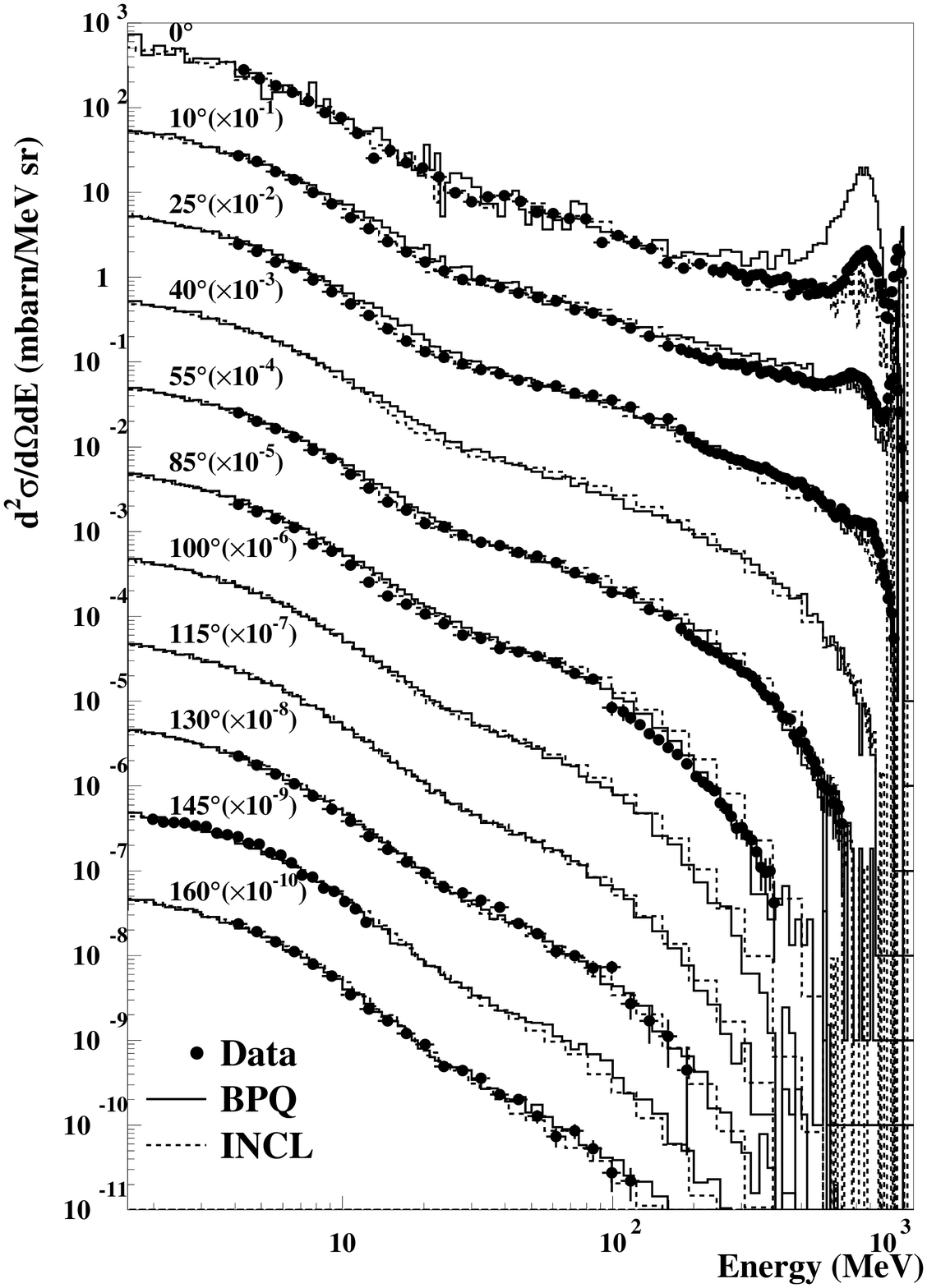}
	\includegraphics[width=7.3cm]{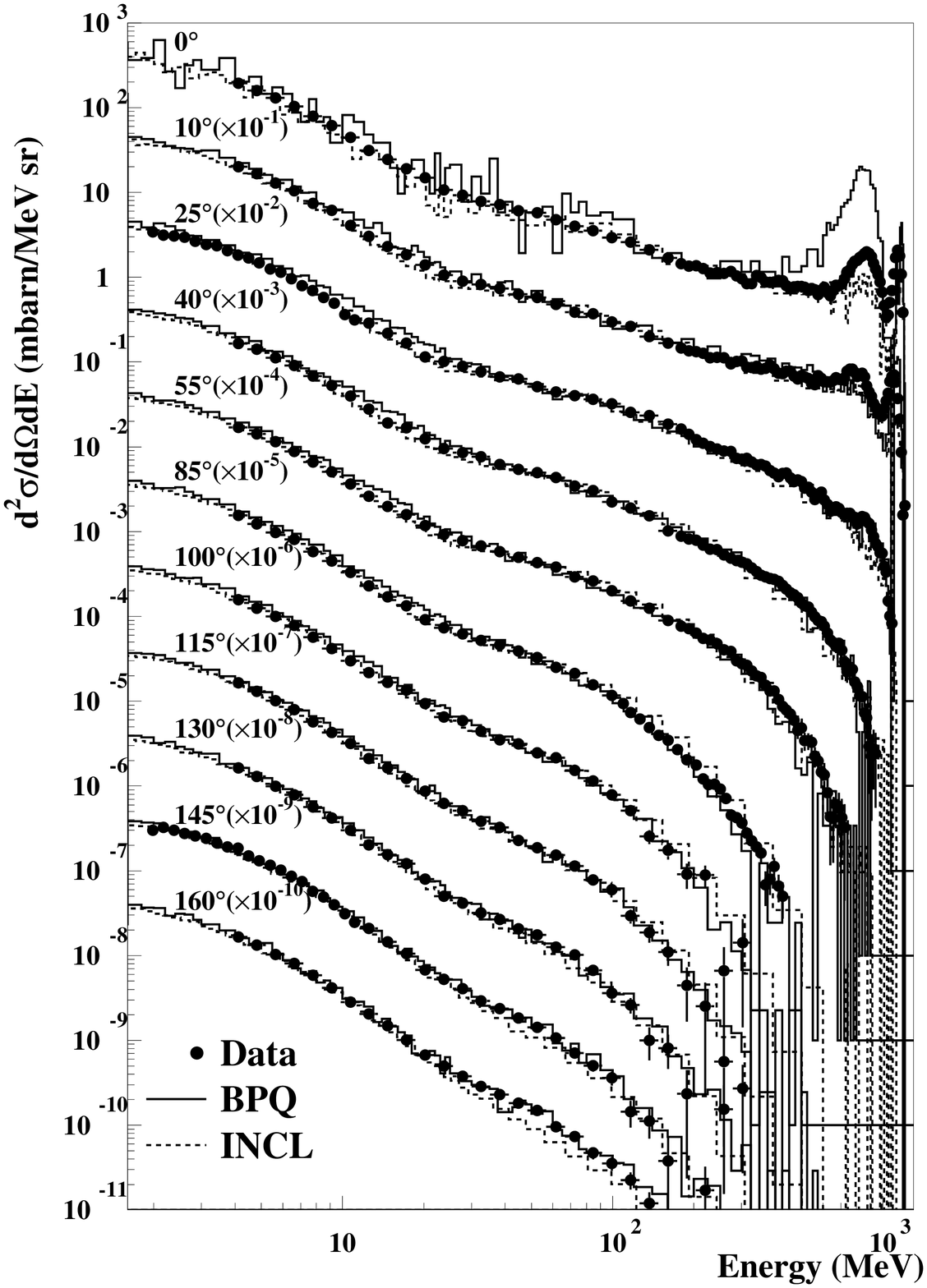}
	\end{center}
	\caption{Experimental p (1200 MeV) + Th (left) and Pb (right) neutron double-differential cross-sections compared with calculations performed with LAHET using either Bertini plus pre-equilibrium ({\it solid line}) or INCL ({\it dashed line}) intra-nuclear cascade model.}
	\label{fig:thpb1200}
	\end{minipage}
\end{figure}

\begin{figure}[htb]
	\begin{minipage}[c]{15.6cm}
	\begin{center}
	\includegraphics[width=7.3cm]{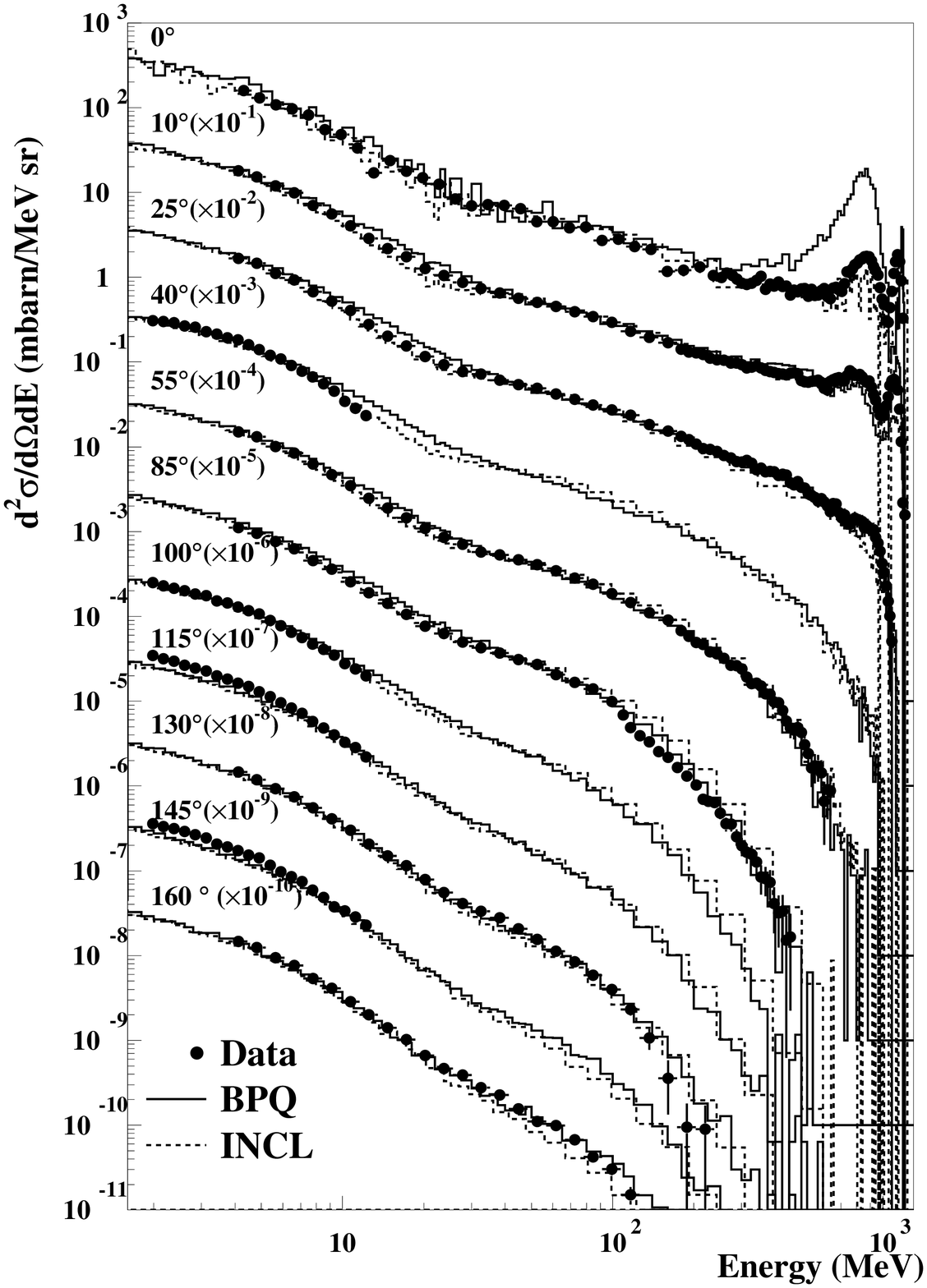}
	\includegraphics[width=7.3cm]{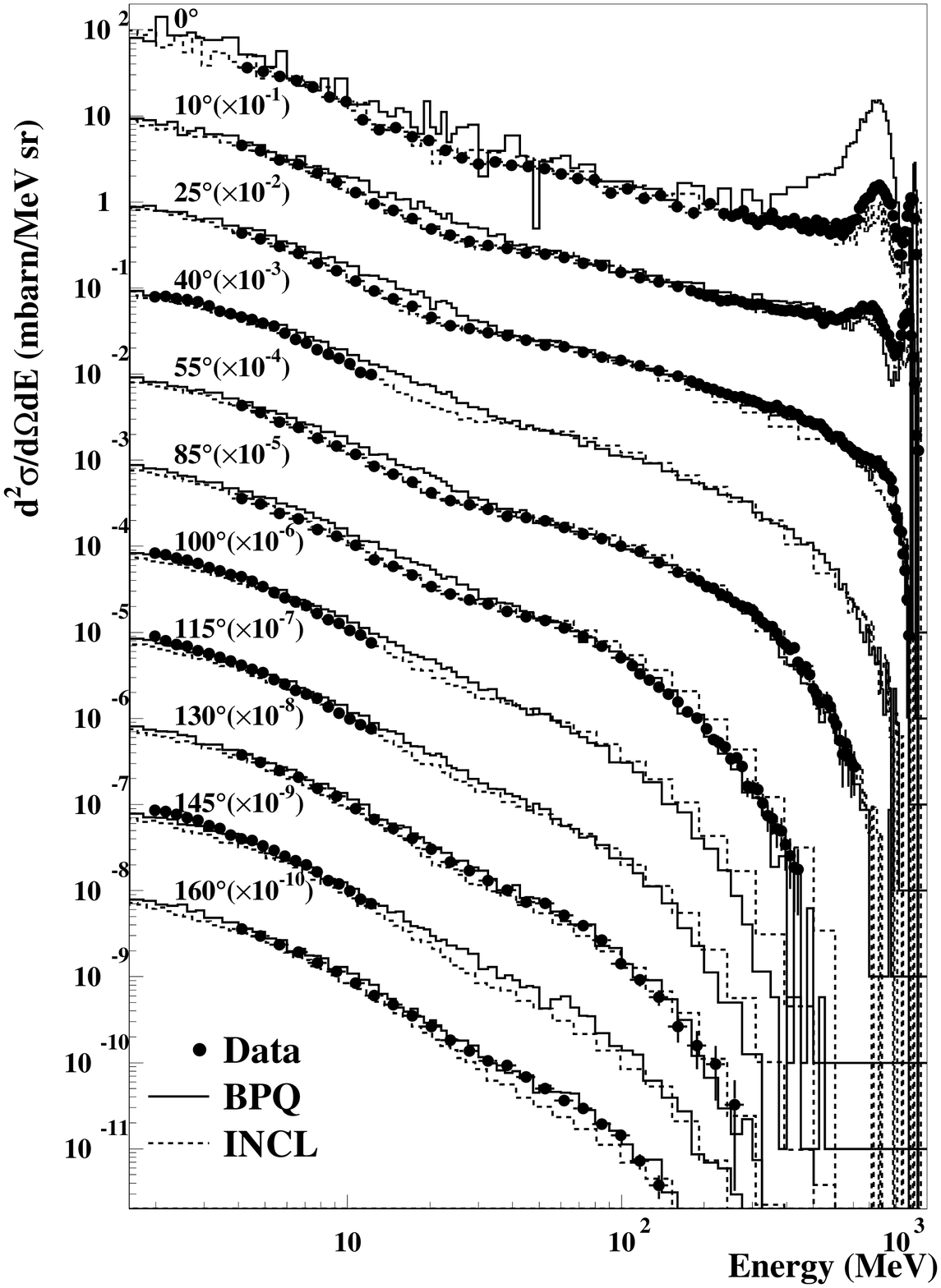}
	\end{center}
	\caption{Experimental p (1200 MeV) + W (left) and Zr (right) neutron double-differential cross-sections compared with calculations performed with LAHET using either Bertini plus pre-equilibrium ({\it solid line}) or INCL ({\it dashed line}) intra-nuclear cascade model.}
	\label{fig:wzr1200}
	\end{minipage}
\end{figure}

\begin{figure}[htb]
	\begin{minipage}[c]{15.6cm}
	\begin{center}
	\includegraphics[width=7.3cm]{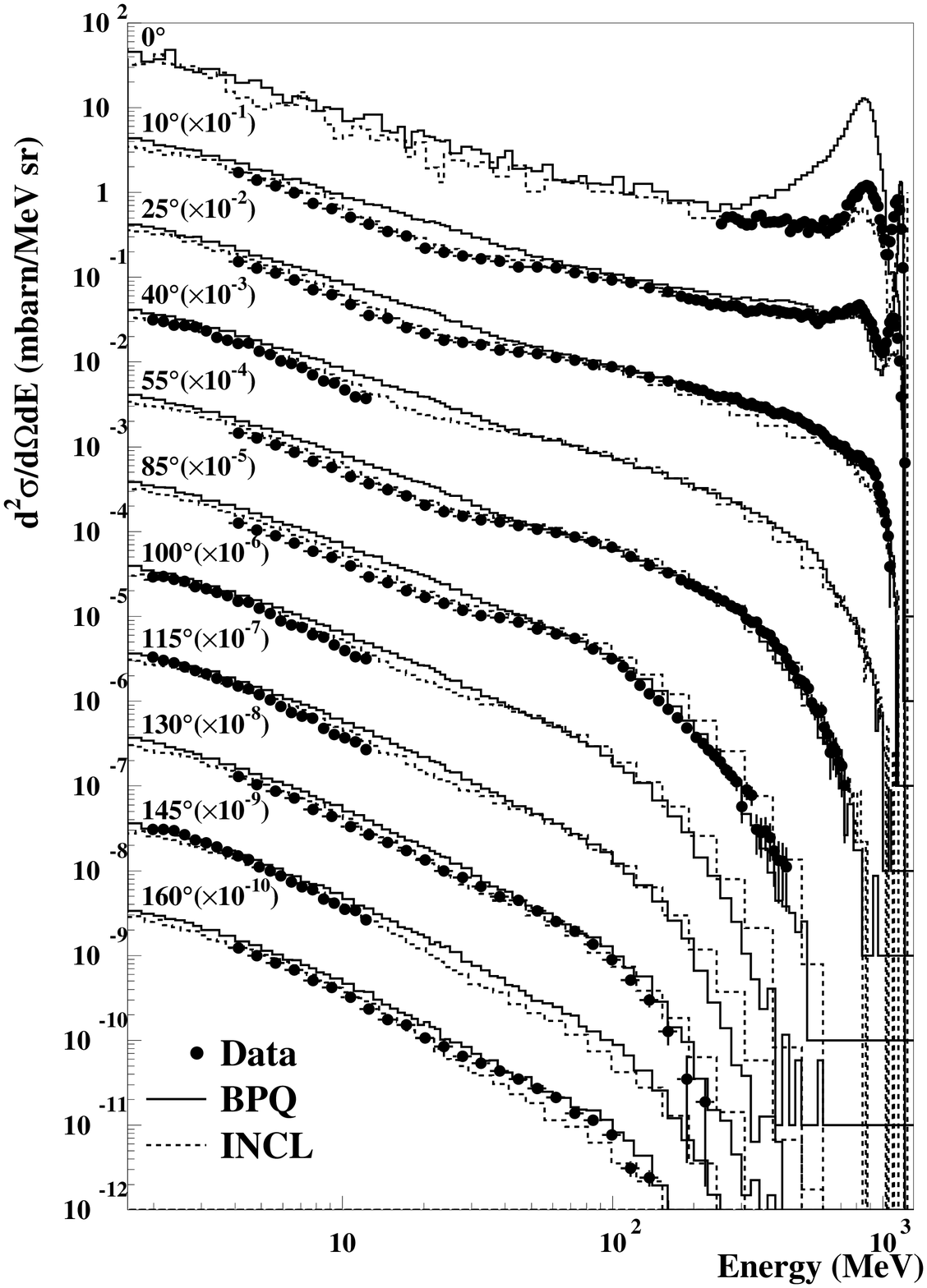}
	\includegraphics[width=7.3cm]{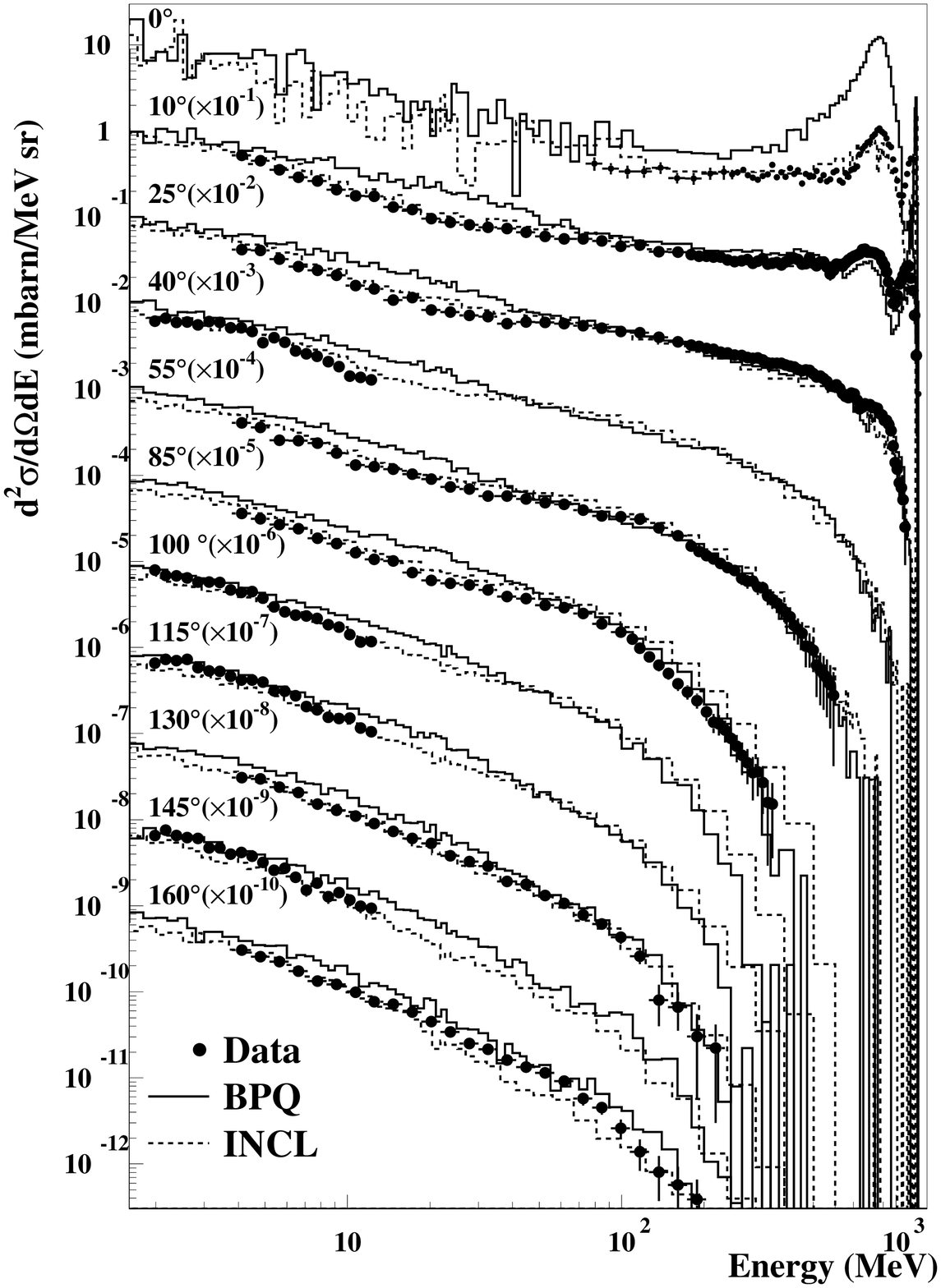}
	\end{center}
	\caption{Experimental p (1200 MeV) + Fe (left) and Al (right) neutron double-differential 
cross-sections compared with calculations performed with LAHET using either Bertini plus pre-equilibrium ({\it solid line}) or INCL ({\it dashed line}) intra-nuclear cascade model.}
	\label{fig:feal1200}
	\end{minipage}
\end{figure}

\begin{figure}[hbt]
	\begin{minipage}[c]{15.6cm}
	\begin{center}
	\includegraphics[width=7.3cm]{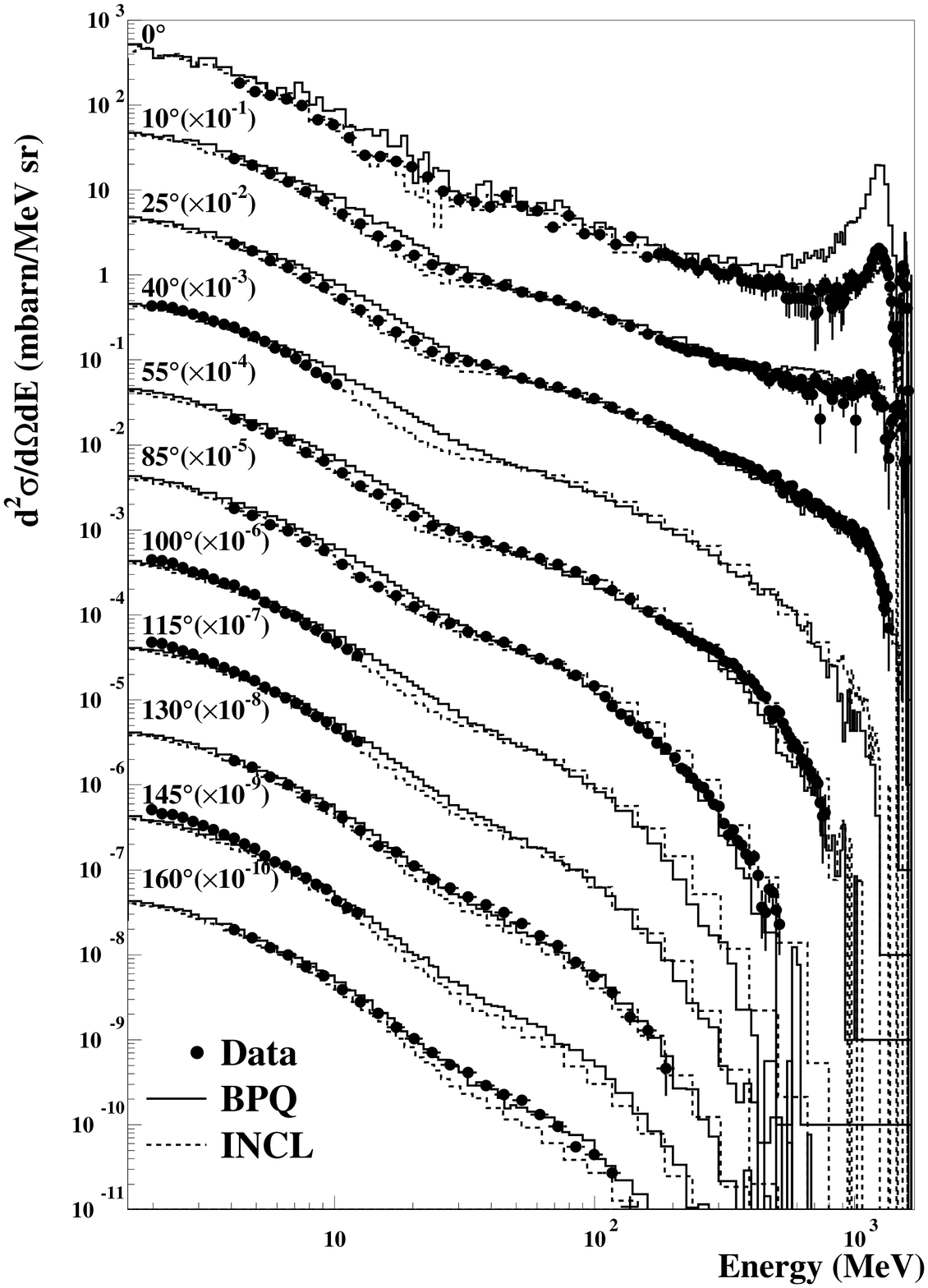}
	\includegraphics[width=7.3cm]{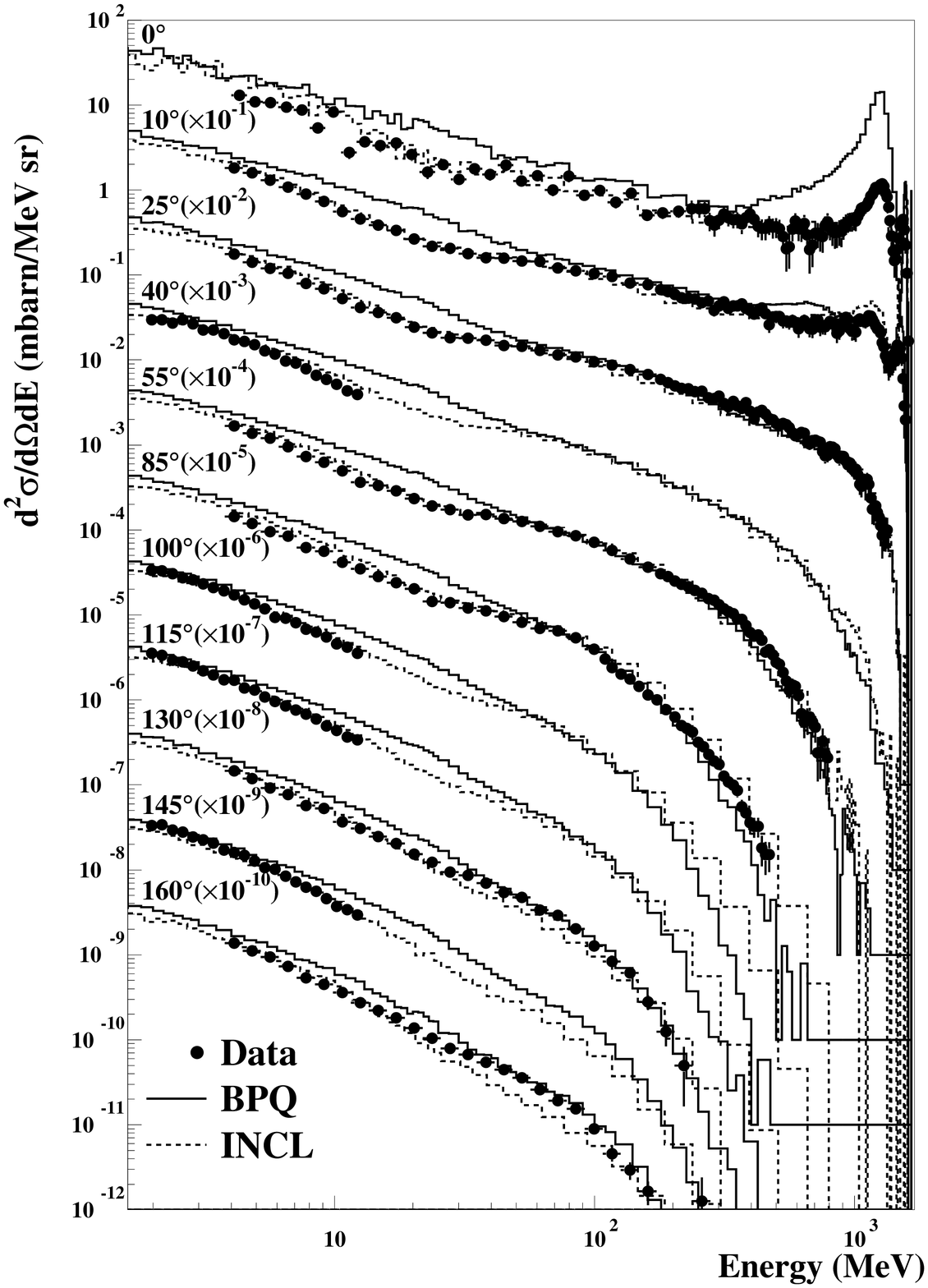}
	\end{center}
	\caption{Id. but for p (1600 MeV) + Pb (left) and Fe (right)}
	\label{fig:pbfe1600}
	\end{minipage}
\end{figure}

At 1200 MeV, the use of Isabel in LAHET being limited to 1 GeV, we compare the data with only BPQ and INCL calculations in figs.~\ref{fig:thpb1200}, \ref{fig:wzr1200} and~\ref{fig:feal1200}, for all the targets. We also performed calculations, which are not shown here, using Bertini without pre-quilibrium. Whatever the target, this model yields too many low energy neutrons, emphasizing that it leads to too high excitation energies. For Th, Pb and W, both BPQ and INCL models give a reasonable agreement with the data, although BPQ tends to slightly overestimate the production of intermediate energies neutrons. As the 
target becomes lighter, this trend is amplified and BPQ begins to also overpredict low energy 
cross-sections. This is an indication that the addition of a pre-equilibrium stage after intra-nuclear cascade to decrease the too large excitation energy found in Bertini may not be the proper solution: in fact, it seems difficult to obtain the correct evaporation neutron production without overestimating intermediate energy cross-sections (which are enhanced by pre-equilibrium). On the contrary, INCL reproduces quite well the results for all the targets, proving that the model has a correct mass dependence. Only for the light targets at very backward angles, the high energy neutron production is underpredicted.

At 1600 MeV, fig.~\ref{fig:pbfe1600} displays the results for the Pb and Fe targets. For BPQ the tendencies noticed at 1200 MeV are growing worse: even for Pb, the agreement is not very good between 10 and 40 MeV. Since the high energy part of the spectra is always rather well reproduced (except at 0$^{\circ}$), this seems to point out a wrong dependence of the pre-equilibrium emission also with incident energy. Here again, INCL gives a satisfactory agreement with the data for both targets. 

In summary, we can conclude that INCL is able to globally reproduce the bulk of our data, with some slight discrepancies in the angular distributions. The Bertini model followed by pre-equilibrium, although it is found to be an improvement compared to Bertini alone, works well for Pb at 800 MeV but fails as the energy is increased and the target gets lighter.  
 
\section{Average multiplicities per reaction neutron}\

\begin{figure}[hbt]
\begin{minipage}[c]{16.2cm}
\begin{center}
\includegraphics[width=8.0cm]{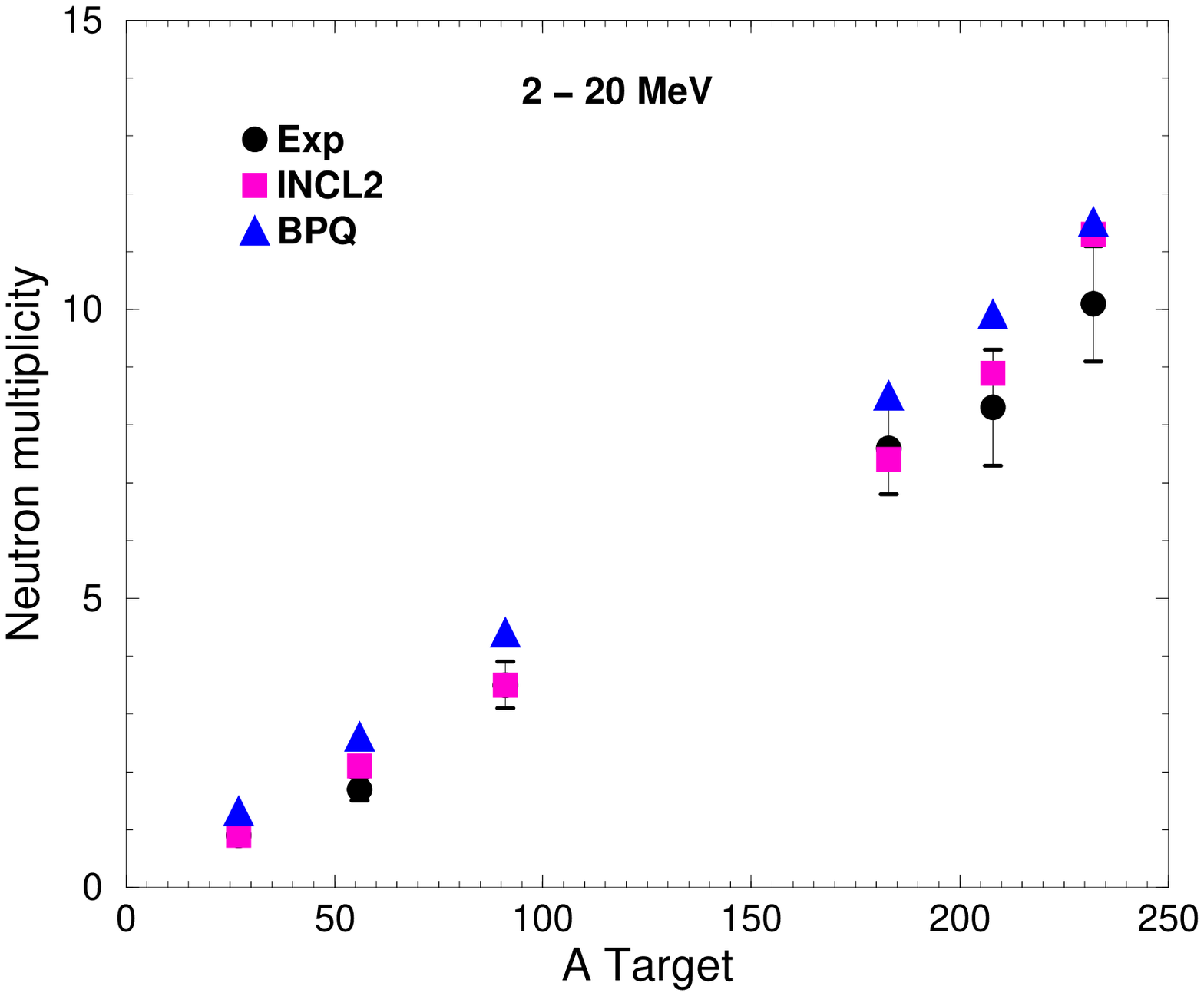}
\includegraphics[width=8.0cm]{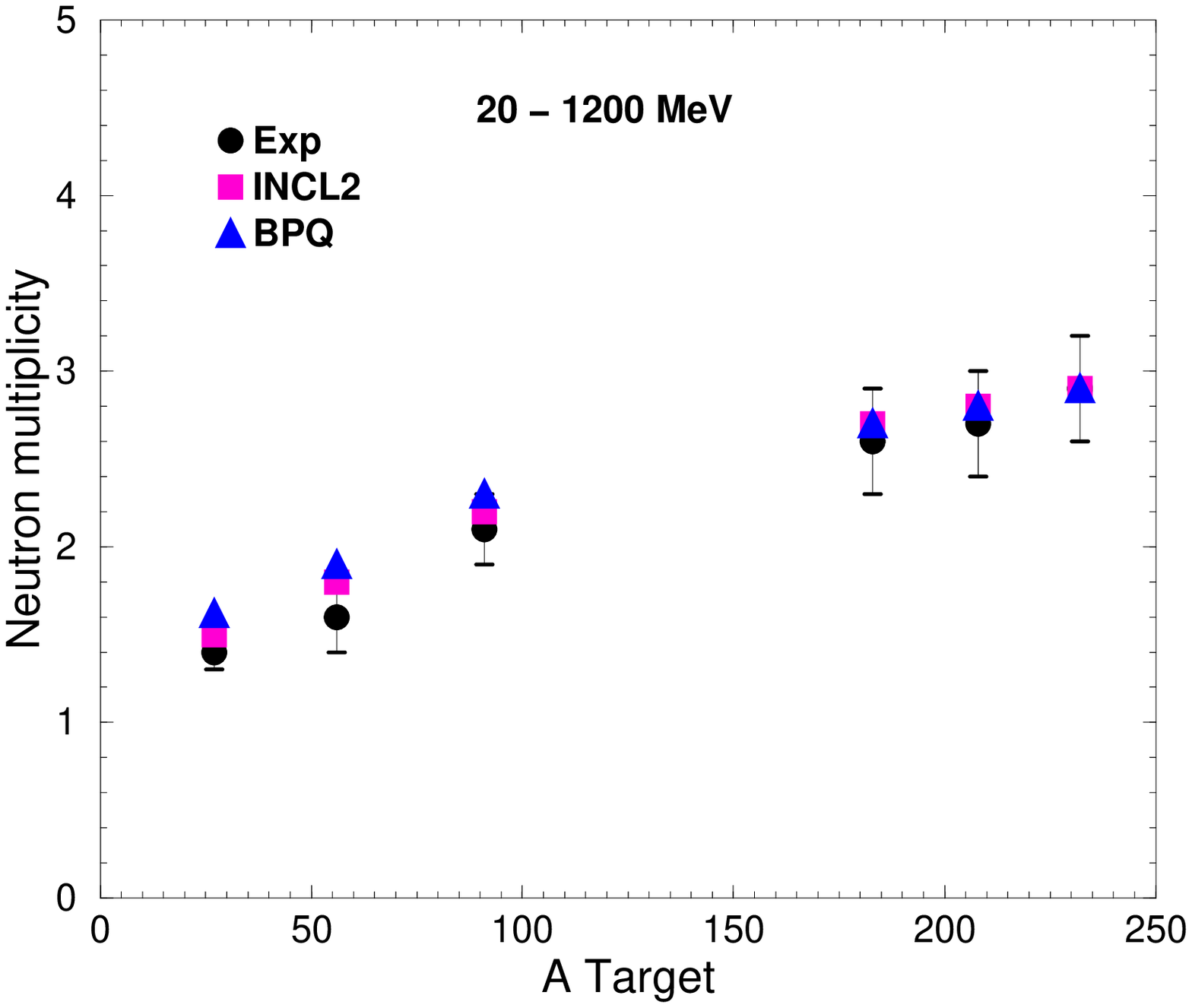}
\end{center}
\caption{Average neutron multiplicities per primary reaction at 1200 MeV for the different targets. Left: 
[2-20 MeV] neutron multiplicities; Right: [20 MeV-Einc] neutron multiplicities.}
\label{fig:mul1200}
\end{minipage}
\end{figure}

\begin{figure}[hbt]
\begin{minipage}[c]{16.2cm}
\begin{center}
\includegraphics[width=8.0cm]{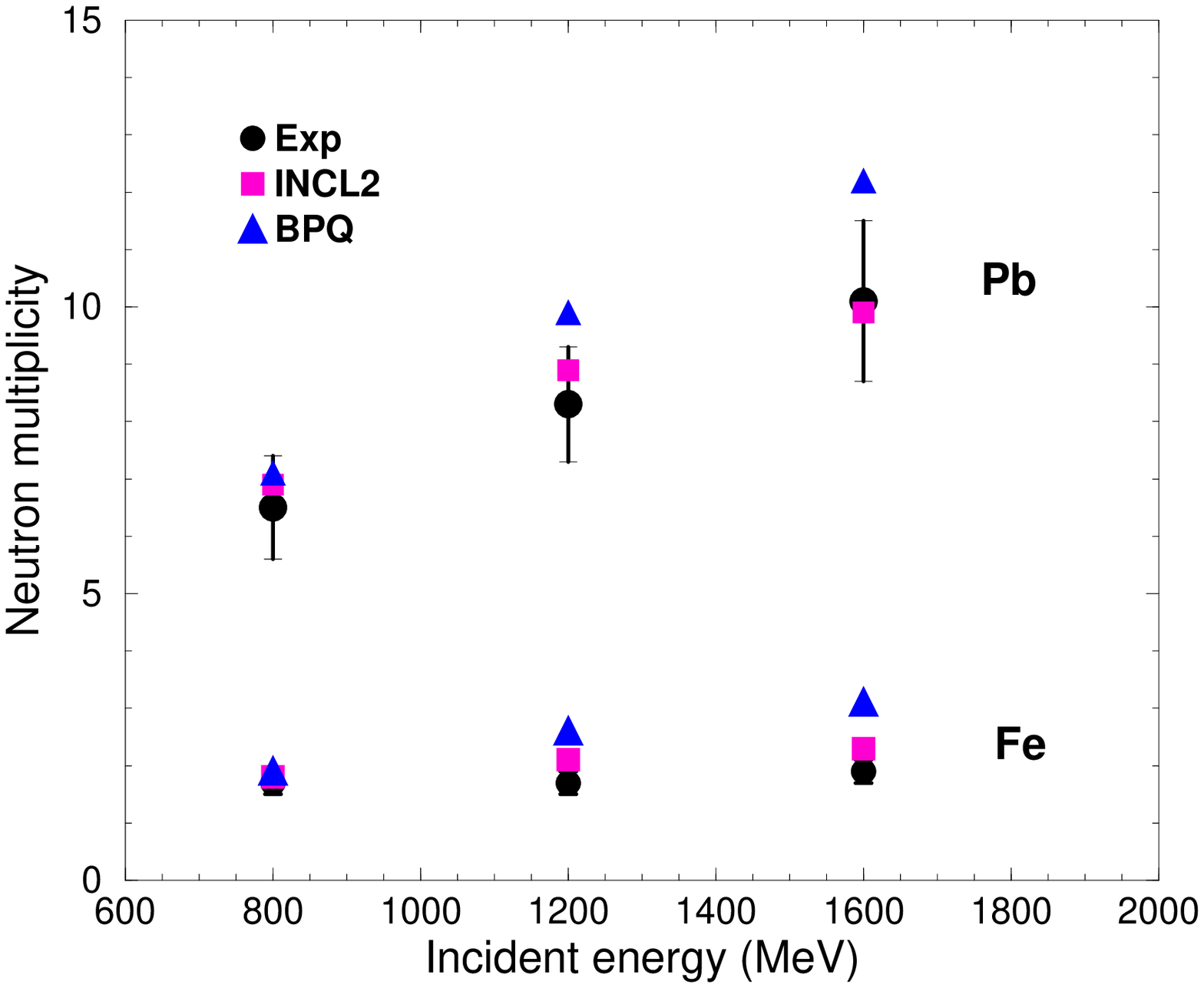}
\includegraphics[width=8.0cm]{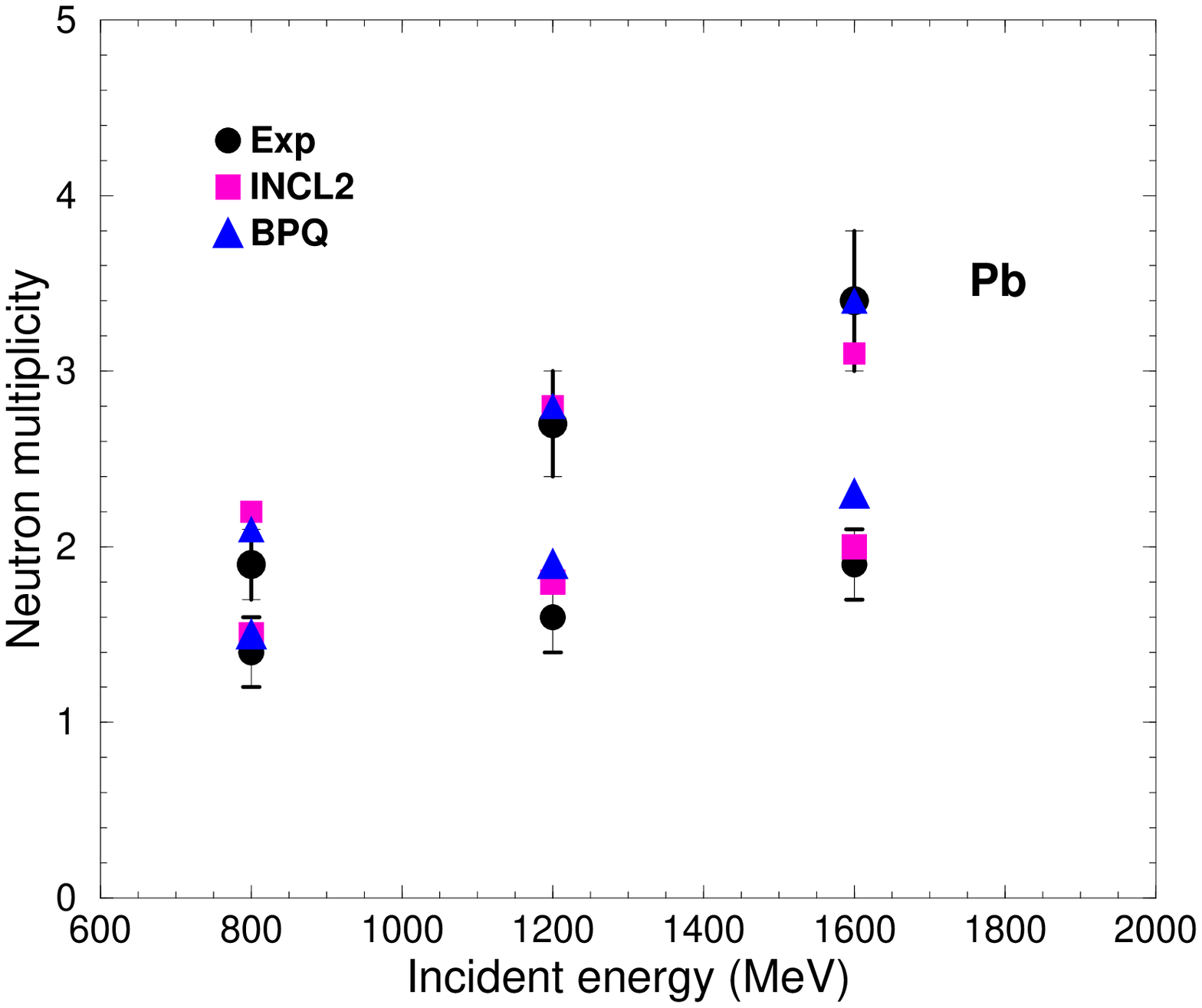}
\end{center}
\caption{Average neutron multiplicities per primary reaction for Pb and Fe as a function of incident 
energy. Left: [2-20 MeV] neutron multiplicities; Right: [20 MeV-Einc] neutron multiplicities.}
\label{fig:mulene}
\end{minipage}
\end{figure}

Since our double-differential cross-sections nearly cover the full angular range with sufficiently close measurements, it has been possible to infer from the data average neutron multiplicities per reaction above the energy threshold of our detectors. This has been done by interpolating between the measured angles, integrating over 4${\pi}$ and then dividing by the reaction cross-section taken from~\cite{Bar}) and the result is shown in tables~\ref{table:multi800}, \ref{table:multi1200} and \ref{table:multi1600} for 2 different energy bins, corresponding roughly to evaporation and cascade neutrons respectively. Since we divide by the reaction cross-section, the multiplicities obtained are numbers of neutrons per primary reaction and therefore contain the effect of secondary reactions. The interpolation between angles was done assuming that the angular dependence of the cross-sections was the same as the one calculated by the TIERCE code. This was necessary in particular between 2 and 4 MeV where we have only a few points from the DENSE detectors. The uncertainty on these interpolations was assessed by using different intra-nuclear cascade models in TIERCE and different interpolation procedures. The values given in the tables take into account this uncertainty plus the systematic errors discussed in section 2. The experimental values are compared with the neutron multiplicities given by the two codes, TIERCE-INCL and LAHET-BPQ. For the calculations, 0-2 MeV and total multiplicities are also given. Results are also shown in fig.~\ref{fig:mul1200} as a function of the mass of the targets studied at 1200 MeV and in fig.~\ref{fig:mulene} versus incident energy for the iron and lead targets. The comparison of the experimental to calculated average neutron multiplicities confirms in a rather concise way what has been observed in the preceeding section. In all cases, INCL agrees with the data within the error bar while BPQ tends to overpredict 2-20 MeV neutron multiplicities, i.e. evaporation neutron production, especially at 1200 and 1600 MeV. For high energy neutrons (above 20 MeV) the sensitivity to the models is less important. This is most likely because of compensating effects, BPQ predicting more intermediate energy neutrons because of pre-equilibrium while INCL spectra often extend to higher energies. However, a significant deviation from the experiment is found at the two highest energies for iron with BPQ.

\begin{figure}[hbt]
	\includegraphics[width=6.6cm]{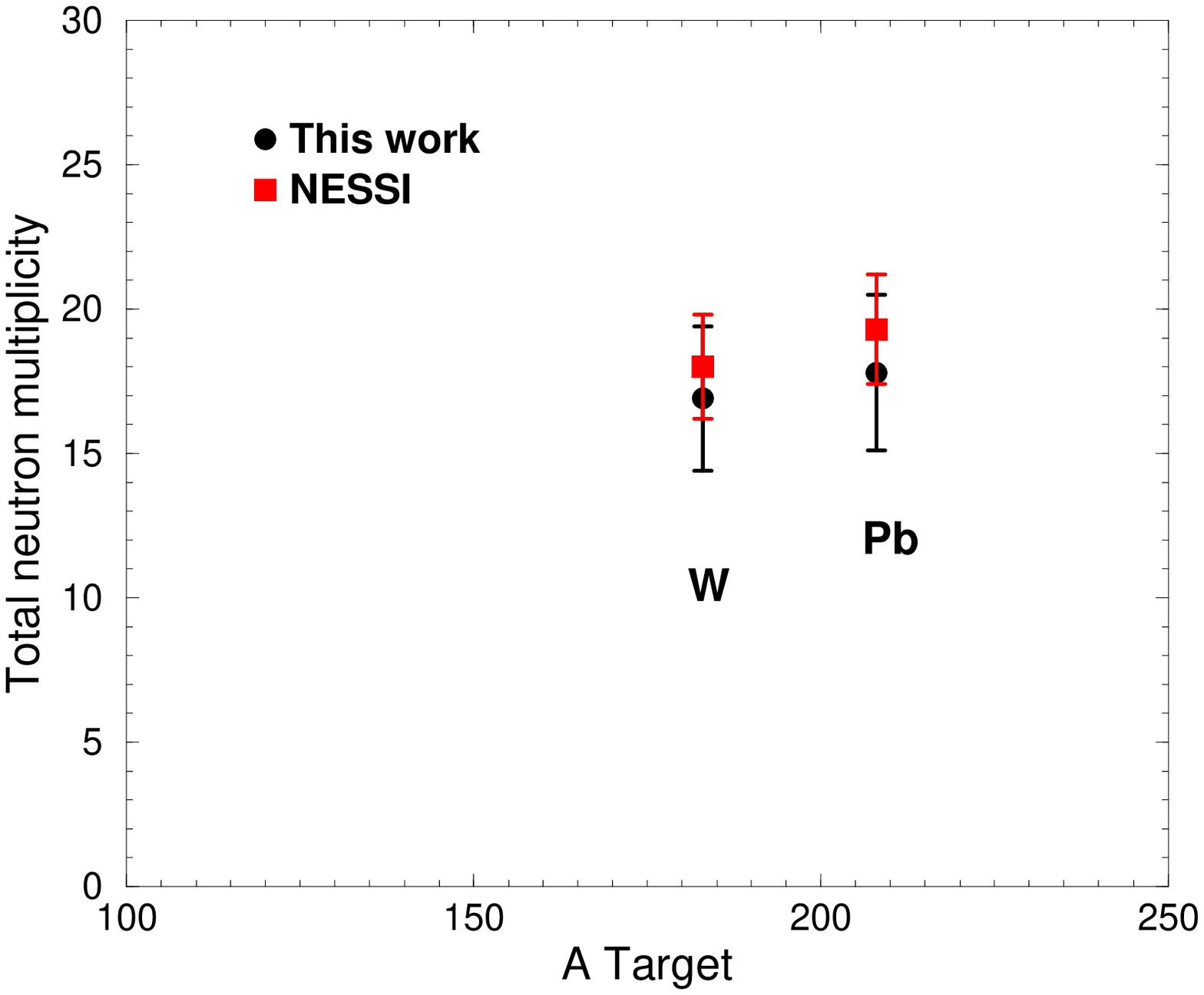}
	\caption{Total neutron multiplicities per primary reaction at 1200 MeV for 15cm diameter W (1cm thick) and Pb (2cm thick) targets obtained by the NESSI collaboration after efficiency and secundary reactions in the liquid scintillator corrections~\protect\cite{AL,DF} compared to the values estimated from our measurement.}
	\label{fig:nessi}
\end{figure}

Neutron multiplicity distributions have been measured by the NESSI collaboration~\cite{AL,DF} using the technique of a gadolinium loaded liquid scintillator tank. The efficiency of this type of detector is large for low energy neutrons and decreases rapidly above 20 MeV. Mean neutron multiplicities were obtained by NESSI at 1200 MeV on W and Pb targets with the same thickness as ours. Although the threshold of our detectors did not allow us to measure neutrons with energies lower than 2 MeV, it is tempting to compare the two results. For Pb at 1200 MeV, the NESSI collaboration measured for a 2cm thick, 15cm diameter target, 14.6 neutrons per reaction which after efficiency correction~\cite{AL,DF} amount to 20.3. In \cite{DF} it is indicated that the average multiplicities have not been corrected for the additional neutrons coming from secundary reactions in the liquid scintillator. This effect was however investigated and estimated to be 5\% for the 2cm thick Pb target (see fig.8 of \cite{DF}). Taking this correction into account, the NESSI multiplicity is then 19.3. In our case, we can estimate the total number of neutrons by adding the experimental values measured between 2 and 1200 MeV to the [0-2 MeV] multiplicity given by the codes. If we take the average between INCL and BPQ values we find 16.9 neutrons per reaction. The error is estimated to be of the order of 15\% taking into account the errors discussed above plus the extrapolation to low energies. While the thickness of the targets in both experiments is the same the diameter is somewhat larger in the NESSI case. We performed a simulation with LAHET to investigate this effect. It appeared that because of secondary reactions of particles emitted sidewards the number of neutrons is 5\% larger than with a 3cm diameter target. This means that for a 15cm diameter target we would find 17.8 $\pm$ 2.7 neutrons to be compared to 19.3 $\pm$ 1.9 if we take an uncertainty of 10\% for the NESSI result. The same can be done for the 1cm thick W target. In this case, the additional contribution due to the target diameter is 3\%. Therefore, we obtain 16.9 $\pm$ 2.5 to be compared to 18.0 $\pm$ 1.0 for NESSI~\cite{AL} after efficiency correction and if we assume, as in Pb, 5\% additional neutron from secundary reactions in the liquid scintillator. As it can be seen in fig.~\ref{fig:nessi}, we can conclude that our results are compatible with those from NESSI within the error bars.   

Also shown in tables~\ref{table:multi800}, \ref{table:multi1200}, \ref{table:multi1600} are the averaged kinetic energies carried away by the neutrons, extracted from the double-differential cross-sections multiplied by energy using the same procedure as for the multiplicities and compared with the calculations. For the 2-20 MeV bin, conclusions similar to what was stated for multiplicities can be drawn, reflecting the fact that Ex$M_n$ is governed by $M_n$ in so far as the same evaporation model is used in both calculations and thus gives an identical energy spectrum for the low energy neutrons. For the high energy bin, the compensating effect noticed for the multiplicities seems to be even stronger and, regarding our uncertainties, it is not possible to discriminate between the two models. It is interesting, nevertheless, to remark that these high energy neutrons carry out the major part (from 80\% for Th to 98\% for Al) of the emitted neutron energy and a large amount (about 30\%) of the incident proton energy. In a thick target this will play an important role in the spatial distribution of the energy deposition and particle production.

\begin{table}[!h]
\begin{center}
\small
\begin{tabular}{cccc||ccc}
\small
{\bf {Energies}}& $\mathbf {M_n^{exp}}$ & $\mathbf {M_n^{INCL}}$ & $\mathbf {M_n^{BPQ}}$ & $\mathbf {E 
\times 
M_n^{exp}}$ & $\mathbf {E \times M_n^{INCL}}$ & $\mathbf {E \times M_n^{BPQ}}$\\
\hline\hline
\multicolumn{7}{c} {$\mathbf {E_{inc.}}$ = {\large {{\bf {800 MeV}}}}}\\
\hline\hline
\multicolumn{7}{c} {{\bf {Pb(p,xn)X}} \hspace{4.5cm} $\sigma_R$ = 1723 mb}\\
\hline
0-2 MeV 	&		&    4.9     &  5.2 	 &		&     5. 	& 5.	\\
2-20 MeV 	& 6.5 $\pm$ 0.7 &    6.9     &  7.1 	 &  38. $\pm$ 4.&    42. 	& 42.	\\
20-$E_{inc.}$	& 1.9 $\pm$ 0.2 &    2.2     &  2.1 	 &200. $\pm$ 20.&   211. 	& 224.	\\
{\bf {Total}}	&		&{\bf {14.0}}&{\bf {14.4}}&		&{\bf {258.}}	&{\bf {271.}}\\
\hline
\multicolumn{7}{c} {{\bf {Fe(p,xn)X}} \hspace{4.5cm} $\sigma_R$ = 776 mb}\\
\hline
0-2 MeV 	&		&    1.0     & 1.3 	 & 		& 1. 		& 1.	\\
2-20 MeV 	& 1.7 $\pm$ 0.2 &    1.8     & 1.9  	 & 12. $\pm$ 1. & 13. 		& 14.	\\
20-$E_{inc.}$	& 1.4 $\pm$ 0.1	&    1.5     & 1.5	 &188. $\pm$ 19.& 175. 		& 203.	\\
{\bf {Total}}	&		&{\bf {4.3}} &{\bf {4.7}} &		&{\bf {189.}}	&{\bf {218.}}\\
\hline\hline
\end{tabular}
\caption{Average neutron multiplicities per primary reaction and kinetic energy carried out by the neutrons (MeV) obtained by integration of the double-differential cross-sections at 800 MeV and compared with calculations using TIERCE-Cugnon or LAHET-BPQ codes.}
\label{table:multi800}
\end{center}
\end{table}

\begin{table}[!h]
\begin{center}
\small
\begin{tabular}{cccc||ccc}
{\bf {Energy}}& $\mathbf {M_n^{exp}}$ & $\mathbf {M_n^{INCL}}$ & $\mathbf {M_n^{BPQ}}$ & $\mathbf {E 
\times 
M_n^{exp}}$ & $\mathbf {E \times M_n^{INCL}}$ & $\mathbf {E \times M_n^{BPQ}}$\\
\hline\hline
\multicolumn{7}{c} {$\mathbf {E_{inc.}}$ = {\large {{\bf {1200 MeV}}}}}\\
\hline\hline
\multicolumn{7}{c} {{\bf {Th(p,xn)X}} \hspace{4.5cm} $\sigma_R$ = 1837 mb}\\
\hline
0-2 MeV 	&		&    7.2     &   7.9	  &		  &          7. &   7.	\\
2-20 MeV 	&10.1 $\pm$ 1.0 &   11.3     & 	11.5 	  & 62. $\pm$ 6.  &         69. &  72. 	\\
20-$E_{inc.}$	&  2.7$\pm$ 0.3	&    2.9     &   2.9	  &301. $\pm$ 30. &        318. & 324. 	\\
{\bf {Total}}	&		&{\bf {21.4}}&{\bf {22.3}}&		  & {\bf {394.}}&{\bf {403.}}\\
\hline
\multicolumn{7}{c} {{\bf {Pb(p,xn)X}} \hspace{4.5cm} $\sigma_R$ = 1719 mb}\\
\hline
0-2 MeV 	&		&    5.8     &   6.0	  &		  &          6. &   6.	\\
2-20 MeV 	& 8.3 $\pm$ 0.8 &    8.9     &   9.9 	  &  52. $\pm$ 5. &         54. &  62.	\\
20-$E_{inc.}$	& 2.7 $\pm$ 0.3	&    2.8     &   2.8	  & 318. $\pm$ 32.&        309. & 326.	\\
{\bf {Total}}	&		&{\bf {17.4}}&{\bf {18.7}}&	 	  &{\bf {369.}} &{\bf {394.}}\\
\hline
\multicolumn{7}{c} {{\bf  {W(p,xn)X}} \hspace{4.5cm} $\sigma_R$ = 1599 mb}\\
\hline
0-2 MeV 	&		&    5.8     &   6.6	  &		  &          5. &   6.	\\
2-20 MeV 	& 7.6 $\pm$ 0.8 &    7.4     &   8.5	  &  49. $\pm$ 5. &         47. &  55. 	\\
20-$E_{inc.}$	& 2.6 $\pm$ 0.3	&    2.7     &   2.7	  & 313. $\pm$ 31.&        316. & 324. 	\\
{\bf {Total}}	&		&{\bf {15.9}}&{\bf {17.8}}&		  &{\bf {368.}}	&{\bf {384.}}\\
\hline\hline
\multicolumn{7}{c} {{\bf {Zr(p,xn)X}} \hspace{4.5cm} $\sigma_R$ = 1047 mb}\\
\hline
0-2 MeV 	&		&    1.9     &   2.3	  &		  &          2. &  2.	\\
2-20 MeV 	& 3.5 $\pm$ 0.4 &    3.5     &   4.4  	  &  24. $\pm$ 2. &         23. & 31.	\\
20-$E_{inc.}$	& 2.1 $\pm$ 0.2	&    2.2     &   2.3	  & 310. $\pm$ 31.&        300. & 317. 	\\
{\bf {Total}}	&		&{\bf {7.6}} &{\bf {8.9}} &		  & {\bf {325.}}& {\bf {350.}}	\\
\hline
\multicolumn{7}{c} {{\bf {Fe(p,xn)X}} \hspace{4.5cm} $\sigma_R$ = 777 mb}\\
\hline
0-2 MeV 	&		&    1.1     & 	1.5 	  &		  &          1. & 1.	\\
2-20 MeV 	& 1.7 $\pm$ 0.2 &    2.1     & 	2.6        &  13. $\pm$ 1. &         15. & 19.	\\
20-$E_{inc.}$	& 1.6 $\pm$ 0.2	&    1.8     & 	1.9 	  & 275. $\pm$ 26.&        270. & 301.	\\
{\bf {Total}}	&		&{\bf {5.0}} & {\bf {6.0}}&		  & {\bf {286.}}&{\bf {321.}}	\\
\hline
\multicolumn{7}{c} {{\bf {Al(p,xn)X}} \hspace{4.5cm} $\sigma_R$ = 475 mb}\\
\hline
0-2 MeV 	&		&    0.3     &  0.5   	  &		  &        0.4 	& 0.5	\\
2-20 MeV 	& 0.9 $\pm$ 0.1 &    0.9     &  1.3   	  &   7. $\pm$ 1. &        7. 	& 11.	\\
20-$E_{inc.}$	& 1.4 $\pm$ 0.1	&    1.5     &  1.6 	  & 298. $\pm$ 30.&      281. 	& 313.	\\
{\bf {Total}}	&		&{\bf {2.7}} &{\bf {3.4}} &		  &{\bf {287.}} & 325.	\\
\hline\hline
\end{tabular}
\caption{Average neutron multiplicities per primary reaction and kinetic energy carried out by the neutrons (MeV) obtained by integration of the double-differential cross-sections at 1200 MeV and compared with calculations using TIERCE-Cugnon or LAHET-BPQ codes.}
\label{table:multi1200}
\end{center}
\end{table}

\begin{table}[!h]
\begin{center}
\small
\begin{tabular}{cccc||ccc}
{\bf {Energy}}& $\mathbf {M_n^{exp}}$ & $\mathbf {M_n^{INCL}}$ & $\mathbf {M_n^{BPQ}}$ & $\mathbf {E 
\times 
M_n^{exp}}$ & $\mathbf {E \times M_n^{INCL}}$ & $\mathbf {E \times M_n^{BPQ}}$\\
\hline\hline
\multicolumn{7}{c} {$\mathbf {E_{inc.}}$ = {\large {{\bf {1600 MeV}}}}}\\
\hline\hline
\multicolumn{7}{c} {{\bf {Pb(p,xn)X}} \hspace{4.5cm} $\sigma_R$ = 1717 mb}\\
\hline
0-2 MeV 	&		&   6.0      &   6.6 	  &		  &        6. 	& 7.	\\
2-20 MeV 	&10.1 $\pm$ 1.0 &   9.9      &  12.2	  & 65.$\pm$ 7.   &       61. 	& 81.	\\
20-$E_{inc.}$	& 3.4 $\pm$ 0.5	&   3.1      &   3.4	  & 410. $\pm$ 41.&      422. 	& 416.	\\
{\bf {Total}}	&		&{\bf {19.0}}&{\bf {22.2}}&		  & {\bf {489.}}& {\bf {503.}}	\\
\hline
\multicolumn{7}{c} {{\bf {Fe(p,xn)X}} \hspace{4.5cm} $\sigma_R$ = 774 mb}\\
\hline
0-2 MeV 	&		&   1.2      &     1.6    &		  &        1. 	& 1.	\\
2-20 MeV 	& 1.9 $\pm$ 0.2 &   2.3      &     3.1    & 14. $\pm$ 1.  &       16. 	& 24.	\\
20-$E_{inc.}$	& 1.9 $\pm$ 0.3	&   2.0      &     2.3 	  & 341. $\pm$ 34.&      363. 	& 387.	\\
{\bf {Total}}	&		& {\bf {5.5}}&{\bf {7.0}} &		  & {\bf {380.}}& {\bf {412.}}	\\
\hline
\end{tabular}
\caption{Average neutron multiplicities per primary reaction and kinetic energy carried out by the neutrons (MeV) obtained by integration of the double-differential cross-sections at 1600 MeV and compared with calculations using TIERCE-Cugnon or LAHET-BPQ codes.}
\label{table:multi1600}
\end{center}
\end{table}

\section{Conclusion}

In this paper, we have displayed double-differential cross-sections measured on a wide set of targets and at different energies. This has allowed a comprehensive comparison with some of the high energy models commonly used in high energy transport codes for applications. In particular, we have compared different models describing the first stage of the reaction (intranuclear cascade possibly followed by pre-equilibrium) keeping the same model for the de-excitation process. As we had shown in a previous paper that the Bertini intra-nuclear cascade was leading to too high excitation energies at the end of the first reaction stage, we have tried to use the option, available in LAHET and recommended by its authors, of adding a pre-equilibium stage. This largely improves the predictions of the code. However, discrepancies tend to appear and grow larger as the energy is increased above 800 MeV and as the target becomes lighter. 
The Isabel model was also tried at 800 MeV. It gave reasonable agreement with the lead data but less good one for iron. Unfortunately, since the use of Isabel is limited to 1 GeV in LAHET, it was not possible to test the energy dependence of this model. Finally, we have shown that the use of the Cugnon intranuclear cascade model, INCL, implemented in the TIERCE code, is able to fairly reproduce the whole bulk of our results. However, it should be recalled that this model still suffers from serious deficiencies mostly due to the fact this it does not treat correctly the diffuseness of the nuclear surface. This was the reason why we had to renormalize the calculation to the correct total reaction cross-section. Also, the sharp surface approximation makes it impossible to have a correct prediction of the most peripheral collisions: this was clearly seen, for instance, in the isotopic distributions of residual nuclei close to the projectile presented in ref.~\cite{Wla}. A new version of the Cugnon model is in progress~\cite{INCL4} which is expected to solve this problem. 

All the data presented will be given to the EXFOR data base or are available on request.

\begin{acknowledgments}
This work is partly supported by the HINDAS EU-project, contract FIKW-CT-2000-00031. We would like to thank O.Bersillon for providing us with the TIERCE code and J.Cugnon and C.Volant with the INCL code. We are also grateful to A.Letourneau for fruitful discussions about the NESSI results.
\end{acknowledgments}

\end{document}